\documentclass[aps,prd,onecolumn,groupedaddress,showpacs,nofootinbib,amssymb]{revtex4-2}
\usepackage[dvips]{graphicx}
\usepackage{amssymb}
\usepackage{amsmath}
\usepackage{graphicx,,color}
\usepackage{amsfonts}
\usepackage{bm}
\usepackage{cancel}
\usepackage{comment}
\usepackage{floatflt}
\usepackage{slashed}
\newcommand\be{\begin{equation}}
\newcommand\ee{\end{equation}}

\allowdisplaybreaks[4]

\begin{document}

\title{Implications of a Scalar Field Interacting with the Dark Matter Fluid on the Primordial Gravitational Waves}
\author{V.K. Oikonomou,$^{1,2}$}\email{voikonomou@gapps.auth.gr;v.k.oikonomou1979@gmail.com}
\affiliation{$^{1)}$Department of Physics, Aristotle University of
Thessaloniki, Thessaloniki 54124, Greece \\ $^{2)}$L.N. Gumilyov
Eurasian National University - Astana, 010008, Kazakhstan}
%$^{2)}$ Laboratory for Theoretical Cosmology, International Center
%of Gravity and Cosmos, Tomsk State University of Control Systems
%and Radioelectronics  (TUSUR), 634050 Tomsk, Russia

 \tolerance=5000

\begin{abstract}
We consider the implications of a scalar field interacting with
the dark matter fluid on the energy spectrum of the primordial
gravitational waves. We choose an interaction type which before a
critical matter density $\rho_m^c$ during the reheating era or
early radiation domination era, the scalar field loses energy
transferring it to the dark matter fluid, while after the critical
matter density $\rho_m^c$ the dark matter fluid loses energy
transferring it to the scalar field. The scalar field is assumed
to have an exponential potential and at the critical matter
density with $\rho_m\sim \rho_m^c$, at which point, the
interaction between the scalar and the dark matter fluid is
switched off, we demand that the effective equation of state of
scalar field is described by a matter dominated era. This is
crucial since it affects the behavior of the trajectories and the
fixed points of the two dimensional dynamical system composed by
the dark matter fluid and the scalar field. Specifically the phase
space contains two stable dark matter dominated final attractors
and two unstable stiff era dominated fixed points. Thus, there
exists the remarkable possibility that the Universe might feel the
passing of the scalar field through the unstable kinetic dominated
fixed points, during the reheating era, with the total equation of
state parameter of the Universe being deformed to be larger than
$w=1/3$. This deformation of the total equation of state parameter
during the reheating era can potentially have significant effects
on the energy spectrum of the primordial gravitational waves. Also
the model we use also contains an $F(R)$ gravity which controls in
a dominant way inflation and the late-time acceleration, in a
phenomenologically viable way.
\end{abstract}

%PACS numbers: 04.50.Kd, 95.36.+x, 98.80.-k, 98.80.Cq

\maketitle

\section{Introduction}

The foundation of all sciences, physics, is currently at a turning
point in its development. The most fundamental aspects of physics
related with the behavior of the Universe at its early times, is
now realistically put to the test. The Large Hadron Collider (LHC)
in CERN has only provided concrete information on the existence of
one Higgs particle, with a low mass, which puts supersymmetric
scenarios into question. Apart from that, currently the
center-of-mass energy at the LHC exceeds 15$\,$TeV, and no new
physics has emerged to date from the LHC. Thus, the burden of
explaining the microcosm falls to the observations coming from the
sky. And now the oxymoron picture emerges in physics, the large
scale evolution of the Universe, can be used to reveal the most
fundamental physics of the cosmos, the particle physics
perspective of the microcosm. One of the most elegant and
theoretical consistent theoretical constructions in particle
cosmology, is the inflationary era proposal
\cite{inflation1,inflation2,inflation3,inflation4}, which as a
theory solves many shortcomings of the standard Big Bang
cosmology. Inflation will be tested by the stage 4 Cosmic
Microwave Background (CMB) experiments
\cite{CMB-S4:2016ple,SimonsObservatory:2019qwx} but also from the
future gravitational wave experiments
\cite{Hild:2010id,Baker:2019nia,Smith:2019wny,Crowder:2005nr,Smith:2016jqs,Seto:2001qf,Kawamura:2020pcg,Bull:2018lat,LISACosmologyWorkingGroup:2022jok}.
In the stage 4 CMB experiments, the $B$-modes in the CMB
polarization modes will be probed directly, while in the future
gravitational wave experiments, the primordial tensor modes will
be probed, which are believed to form a stochastic background with
small or negligible anisotropies. Encouraging data for the
existence of a stochastic gravitational wave background were
provided by the NANOGrav and PTA collaborations in June 29 2023
\cite{NANOGrav:2023gor,Antoniadis:2023ott,Reardon:2023gzh,Xu:2023wog},
which renders this date a monumental date for fundamental physics
and large scale astrophysics. After the announcement of the
existence of a stochastic gravitational wave background, many
works emerged that tried to explain the signal from the
cosmological perspective, see for example
\cite{sunnynew,Oikonomou:2023qfz,Cai:2023dls,Han:2023olf,Guo:2023hyp,Yang:2023aak,Addazi:2023jvg,Li:2023bxy,Niu:2023bsr,Yang:2023qlf,Datta:2023vbs,Du:2023qvj,Salvio:2023ynn,Yi:2023mbm,You:2023rmn,Wang:2023div,Figueroa:2023zhu,Choudhury:2023kam,HosseiniMansoori:2023mqh,Ge:2023rce,Bian:2023dnv,Kawasaki:2023rfx,Yi:2023tdk,An:2023jxf,Zhang:2023nrs,DiBari:2023upq,Jiang:2023qbm,Bhattacharya:2023ysp,Choudhury:2023hfm,Bringmann:2023opz,Choudhury:2023hvf,Choudhury:2023kdb,Huang:2023chx,Jiang:2023gfe,Zhu:2023lbf,Ben-Dayan:2023lwd,Franciolini:2023pbf,Ellis:2023oxs,Liu:2023ymk,Liu:2023pau,Madge:2023cak,Huang:2023zvs,Fu:2023aab,Maji:2023fhv,Gangopadhyay:2023qjr,Wang:2023sij,Wang:2023ost},
and also
\cite{Schwaller:2015tja,Ratzinger:2020koh,Ashoorioon:2022raz,Choudhury:2023vuj,Choudhury:2023jlt,Choudhury:2023rks,Bian:2022qbh}
and furthermore
\cite{Guo:2023hyp,Yang:2023aak,Machado:2018nqk,Regimbau:2022mdu}.
The existence of a stochastic gravitational wave background cannot
be explained by standard single field and conformally related
theories by themselves. What is needed to explain the current and
future stochastic gravitational wave backgrounds in an abnormal
reheating era and beyond, with a broken power-law, combined
\cite{sunnynew,Oikonomou:2023qfz,Benetti:2021uea,Vagnozzi:2020gtf}
with low-reheating temperatures
\cite{sunnynew,Oikonomou:2023qfz,Benetti:2021uea,Vagnozzi:2020gtf}
and a blue-tilted inflationary spectrum
\cite{Kamali:2020drm,Brandenberger:2015kga,Brandenberger:2006pr,
Ashtekar:2011ni,Bojowald:2011iq,Mielczarek:2009vi,Bojowald:2008ik,Calcagni:2020tvw,Koshelev:2020foq,Koshelev:2017tvv,Baumgart:2021ptt}.

However, it is possible that the Universe during reheating and the
subsequent radiation domination era may have disturbances in the
total effective equation of state (EoS) parameter, like stiff eras
for example which may cause deformations of the radiation
dominated EoS to be larger than $w=1/3$ in the range $w=(1/3,1)$.
This stiff era assumption has also been studied in the literature
see
\cite{Co:2021lkc,Gouttenoire:2021jhk,Giovannini:1998bp,Oikonomou:2023bah,Ford:1986sy,Kamionkowski:1990ni,Grin:2007yg,Visinelli:2009kt,Giovannini:1999qj,Giovannini:1999bh,Giovannini:1998bp,Harigaya:2023pmw}.
In this line of research in this paper we provide a fundamental
mechanism of how EoS deformations can occur during the radiation
domination era, well before the BBN, and well before the
matter-radiation equality. Specifically we shall focus on modes
with wavenumbers $k=10^{10}-10^{13}\,$Mpc$^{-1}$, which correspond
to the early radiation or even reheating era. The model is based
on the existence of a scalar field, in the presence of matter and
radiation fluids and in the presence of an $F(R)$ gravity
\cite{reviews1,reviews2,reviews3,reviews4,reviews5,Sebastiani:2016ras,reviews6}.
More importantly we assume the existence of a non-trivial
interaction between the scalar and matter fluids, which before a
critical matter density $\rho_m^c$, during the radiation
domination era, acts in such a way so that the scalar field fluid
loses energy and transfers it to the matter fluid, when
$\rho_m\sim \rho_m^c$ the interaction is zero, and when $\rho_m >
\rho_m^c$ the interaction flips its sign and the dark matter fluid
loses its energy and transfers it to the scalar field. Such
interacting fluids models in cosmology have thoroughly been
investigated in the literature, see for example
\cite{Copeland:1997et,Boehmer:2008av,Yang:2022csz,Nunes:2022bhn,Gariazzo:2021qtg,Nunes:2021zzi,Yang:2021flj,Yang:2019uzo,Yang:2019qza,Yang:2018uae}
and references therein. By construction, the $F(R)$ gravity
dominates the evolution during the inflationary era, and during
the dark energy era. In between the evolution is dominated by
radiation and after the critical matter density era $\rho_m^c$, by
the scalar field and dark matter fluids competing with the
radiation fluid as the Universe evolves. We form the two
dimensional subspace of the phase space of the cosmological system
under study, composed by the scalar field and dark matter fluids
and we calculate the fixed points. As we show, under certain
assumptions, exactly at the critical matter density $\rho_m^c$,
there exist two dark stable dark matter attractors in the phase
space and two kination fixed points which are unstable. As we
show, there exist trajectories in the phase space that end up to
the final dark matter attractors, but before that, these pass
through the stiff era fixed points of the cosmological system.
Thus an exciting possibility emerges, that the Universe
experienced EoS deformations before the BBN era (or simply the
total EoS during radiation might be larger than $w=1/3$), in which
it stayed for a short time. After that, the scalar field reaches
the dark matter attractors, and thus the Universe returned to the
radiation domination evolution again. Accordingly we calculate the
energy spectrum of the primordial gravitational waves including
the short EoS deformations effects, and we show that the predicted
signal can be detectable from the future LISA, SKA, BBO and DECIGO
experiments, but not from the Einstein Telescope. Also we briefly
discuss the issues that may arise with the EoS deformations,
related to the abundances of the light elements and the sound
speed of the CMB modes at the last scattering surface.

\section{Scalar field-dark matter-fluid Interactions and the Possibility of total-EoS Deformations}

\subsection{General Theoretical Framework}

The theoretical framework we shall consider in this section
consists from an $F(R)$ gravity theory in the presence of a scalar
field with exponential potential and in the presence of a dark
matter fluid and a radiation perfect fluid. We also consider a
non-trivial interaction between the dark matter fluid and the
scalar field, which as will be proven, it plays an important role
for the analysis that will follow. The gravitational action of the
model which we shall consider is,
\begin{equation}
\label{mainaction} \mathcal{S}=\int d^4x\sqrt{-g}\left[
\frac{1}{2\kappa^2}F(R)-\frac{1}{2}\partial^{\mu}\phi\partial_{\mu}\phi-V(\phi)+\mathcal{L}_m
\right]\, ,
\end{equation}
with $\kappa^2=\frac{1}{8\pi G}=\frac{1}{M_p^2}$, while $G$
denotes as usual Newton's gravitational constant,  and $M_p$
stands for the reduced Planck mass. The term $\mathcal{L}_m$
contains the matter fluids present, and specifically the dark
matter and radiation fluid. Since we will assume that the dark
matter fluid interacts with the scalar field, only the radiation
fluid is considered to be a perfect fluid. Regarding the $F(R)$
gravity, it will be assumed to have the following form,
\begin{equation}\label{starobinsky}
F(R)=R+\frac{1}{M^2}R^2-\gamma \Lambda
\Big{(}\frac{R}{3m_s^2}\Big{)}^{\delta}\, .
\end{equation}
The $R^2$ term  of the above $F(R)$ gravity will control the
early-time evolution, while the last term will dominate the
late-time evolution synergistically with the matter fluids.  Note
that $m_s$ in Eq. (\ref{starobinsky}) is equal to
$m_s^2=\frac{\kappa^2\rho_m^{(0)}}{3}$, also the parameter $\delta
$ is assumed to be positive and specifically $\delta=1/100$, while
$\gamma=1/0.5$ and $\Lambda $ is the cosmological constant at
present day. Finally the $R^2$-term related parameter $M$, is
chosen to be $M= 1.5\times
10^{-5}\left(\frac{N}{50}\right)^{-1}M_p$ on a pure early time
phenomenological basis \cite{Appleby:2009uf}, where $N$ denotes
the $e$-foldings number. Regarding the scalar field potential
$V(\phi)$, it is assumed to have the following exponential form,
\begin{equation}\label{exponentialpotential}
V(\phi)=V_0\,e^{-\lambda \phi \kappa}\, ,
\end{equation}
where the parameter $V_0$ is assumed to be quite smaller that
$R^2/M^2$, that is $V_0\ll \frac{R^2}{M^2}$, without loss of
generality, and recall $\kappa=1/M_p$. With the assumption of a
flat Friedmann-Robertson-Walker (FRW) geometric background,
\begin{equation}
\label{metricfrw} ds^2 = - dt^2 + a(t)^2 \sum_{i=1,2,3}
\left(dx^i\right)^2\, ,
\end{equation}
the variation of the gravitational action with respect to the
metric and the scalar field yields the following field equations,
\begin{align}\label{eqnsofmkotion}
& 3 H^2F_R=\frac{RF_R-F}{2}-3H\dot{F}_R+\kappa^2\left(
\rho_r+\rho_m+\frac{1}{2}\dot{\phi}^2+V(\phi)\right)\, ,\\ \notag
& -2\dot{H}F=\kappa^2\dot{\phi}^2+\ddot{F}_R-H\dot{F}_R
+\frac{4\kappa^2}{3}\rho_r\, ,
\end{align}
\begin{equation}\label{scalareqnofmotion}
\ddot{\phi}+3H\dot{\phi}+V'(\phi)=\frac{Q}{\dot{\phi}}\, ,
\end{equation}
with $F_R=\frac{\partial F}{\partial R}$, and the ``dot''
indicates differentiation with respect to the cosmic time $t$,
while the ``prime'' denotes differentiation with respect to
$\phi$. Also $Q$ is the interaction term between the matter fluid
and the scalar field. Recall that the dark matter fluid and the
scalar field are not perfect fluids, since they are assumed to
interact non-trivially, and in order to see this explicitly let us
rewrite the field equations in the Einstein-Hilbert form in a FRW
metric, as follows,
\begin{align}\label{flat}
& 3H^2=\kappa^2\rho_{tot}\, ,\\ \notag &
-2\dot{H}=\kappa^2(\rho_{tot}+P_{tot})\, ,
\end{align}
with $\rho_{tot}=\rho_{\phi}+\rho_G+\rho_r+\rho_m$ denoting the
total energy density composed by all the cosmological fluids
present, and $P_{tot}=P_r+P_{\phi}+P_{G}$ is the total pressure.
The cosmological fluids present are the dark matter fluid with
energy density $\rho_m$ and zero pressure, the scalar field fluid,
with its energy density $\rho_{\phi}$ and pressure $P_{\phi}$
being equal to,
\begin{equation}\label{rhoscalarandP}
\rho_{\phi}=\frac{\dot{\phi}^2}{2}+V(\phi)\, ,\,\,\,
P_{\phi}=\frac{\dot{\phi}^2}{2}-V(\phi)\, ,
\end{equation}
the radiation fluid with energy density $\rho_r$ and pressure
$P_r=\frac{\rho_r}{3}$, and the effective geometric fluid with its
energy density $\rho_{G}$ and pressure $P_G$ being equal to,
\begin{equation}\label{degeometricfluid}
\rho_{G}=\frac{F_R R-F}{2}+3H^2(1-F_R)-3H\dot{F}_R\, ,
\end{equation}
\begin{equation}\label{pressuregeometry}
P_G=\ddot{F}_R-H\dot{F}_R+2\dot{H}(F_R-1)-\rho_G\, .
\end{equation}
Note that the geometric fluid quantifies the overall effect of the
$F(R)$ gravity. The geometric and radiation fluids are perfect
fluids, however the dark matter fluid and the scalar field are
not, and this can be seen by the continuity equations,
\begin{align}\label{fluidcontinuityequations}
& \dot{\rho}_m+3H(\rho_m)=-Q\, , \\ \notag &
\dot{\rho}_{\phi}+3H(\rho_{\phi}+P_{\phi})=Q\, ,
\\ \notag &
 \dot{\rho}_r+3H(\rho_r+P_r)=0\, , \\
\notag & \dot{\rho}_G+3H(\rho_G+P_G)=0\, .
\end{align}
However, although the dark matter and scalar field fluids interact
and are not perfect fluids, the total cosmological fluid is
conserved and is a perfect fluid, which has the following
continuity equation,
\begin{align}\label{fluidcontinuityequations}
\dot{\rho}_{tot}+3H(\rho_{tot}+P_{tot})=0\, ,
\end{align}
and can be obtained by simply adding the distinct continuity
equations, and observe that the interaction terms cancel. The
specific form of the interaction term $Q$ and the energy transfer
between the dark matter fluid and the scalar field fluid is of
great importance for the rest of the article, so let us specify
the interaction term at this point and let us discuss the specific
features it implies for the evolution of the Universe. The
interaction term will be assumed to have the following form,
\begin{equation}\label{interactionterm1}
Q=\sqrt{\frac{2}{3}}\kappa \beta \rho_m\dot{\phi}\tanh \left(
\frac{\rho_m}{\xi \rho_m^c}-1 \right)\, ,
\end{equation}
with $\beta$ some positive number and $\rho_m^c$ is a critical
specific matter energy density which we assume to have some value
well below the value of the matter energy density at
matter-radiation equality, at an era belonging to some point well
before the BBN era, during the radiation domination era. Note that
the term $\sim \sqrt{\frac{2}{3}}\kappa \beta \rho_m\dot{\phi}$
has a phenomenological basis for scalar-tensor theories
\cite{Wetterich:1994bg}. The behavior of the term $\sim \tanh
\left( \frac{\rho_m}{\rho_m^c}-1 \right)$ is as follows,
\begin{equation}\label{interactionterm2}
\tanh \left( \frac{\rho_m}{\rho_m^c}-1 \right)=\left\{
\begin{array}{c}
  -1,\,\,\,\mathrm{when}\,\,\,\rho_m\ll \rho_m^c \\
  0,\,\,\,\mathrm{when}\,\,\,\rho_m\sim \rho_m^c  \\
  1,\,\,\,\mathrm{when}\,\,\,\rho_m\gg \rho_m^c \, ,
\end{array}\right.
\end{equation}
Thus the interaction term $Q$ behaves as follows,
\begin{equation}\label{interactionterm3}
Q=\left\{
\begin{array}{c}
  -\sqrt{\frac{2}{3}}\kappa \beta \rho_m\dot{\phi},\,\,\,\mathrm{when}\,\,\,\rho_m\ll \rho_m^c \\
  0,\,\,\,\mathrm{when}\,\,\,\rho_m\sim \rho_m^c  \\
  \sqrt{\frac{2}{3}}\kappa \beta \rho_m\dot{\phi},\,\,\,\mathrm{when}\,\,\,\rho_m\gg \rho_m^c \, ,
\end{array}\right.
\end{equation}
So when $\rho_m\ll \rho_m^c$ primordially, the scalar field fluid
loses its energy and transfers it to the dark matter fluid which
gains energy and this behavior continues during the radiation
domination era when $\rho_m\sim \rho_m^c$, where the interaction
between the dark matter fluid and the scalar field fluid switches
off. After that era, the matter fluid gains energy from the scalar
field. Hence it is apparent that the scalar field loses energy
primordially and transfers it to the dark matter fluid.

\subsection{Dynamics of the Universe During the Inflationary Era}

Now before we focus on the two-dimensional subsystem composed by
the dark matter fluid and the scalar field, which will dominate
the evolution during the end of the radiation domination era and
specifically sometime earlier from the matter-radiation equality
and well beyond that, until the dark energy era commences, let us
focus on the dynamical evolution of the Universe at early and at
late times. We start off with early times, where as we will show,
the $R^2$ gravity dominates the evolution. We assume an
intermediate inflationary scale $H_I=10^{13}$GeV, and we also take
into account the Planck value of the Hubble rate at present day
$H_0$, which is \cite{Planck:2018vyg},
\begin{equation}\label{H0today}
H_0=67.4\pm 0.5 \frac{km}{sec\times Mpc}\, ,
\end{equation}
so $H_0=67.4km/sec/Mpc$ which expressed in natural units is
$H_0=1.37187\times 10^{-33}$eV, therefore $h\simeq 0.67$.
Furthermore, according to the latest Planck data, $h$-scaled dark
matter energy density  $\Omega_c h^2$ is,
\begin{equation}\label{codrdarkmatter}
\Omega_c h^2=0.12\pm 0.001\, .
\end{equation}
In order to have a quantitative idea of the order of magnitude of
the various terms appearing in the field equations during the
inflationary era, let us use the above values in the various terms
in the field equations. For the $F(R)$ gravity of Eq.
(\ref{starobinsky}), the field equations (\ref{eqnsofmkotion})
become,
\begin{equation}\label{friedmanequationinflation}
3H^2\left(1+\frac{2}{M^2}R-\delta \gamma
\Big{(}\frac{R}{3m_s^2}\Big{)}^{\delta-1}\right)=\frac{R^2}{2M}+(\gamma-\gamma
\delta
)\frac{\Big{(}\frac{R}{3m_s^2}\Big{)}^{\delta}}{2}-3H\dot{R}\Big{(}\frac{2}{M^2}-\gamma
\delta (\delta-1)\Big{(}\frac{R}{3m_s^2}\Big{)}^{\delta-2}\Big{)}+
\kappa^2\Big{(}\rho_r+\rho_m+\frac{1}{2}\kappa^2\dot{\phi}^2+V(\phi)
\Big{)}\, .
\end{equation}
Using the values of the free parameters quoted below Eq.
(\ref{starobinsky}), which yield a viable dark energy era
\cite{Oikonomou:2020oex}, we shall compare the various terms in
(\ref{friedmanequationinflation}) in order to find the dominant
terms. Using the inflationary slow-roll assumption $\dot{H}\ll
H^2$, the Ricci scalar during inflation is of the approximately
$R\simeq 12 H^2$, hence for $H= H_I\sim 10^{13}$GeV, the curvature
scalar becomes approximately $R\sim 1.2\times 10^{45}$eV$^2$.
During inflation we also have, $R^2/M^2\sim \mathcal{O}(1.55\times
 10^{45})$eV$^2$, also $\sim \Big{(}\frac{R}{3m_s^2}\Big{)}^{\delta}\sim
 \mathcal{O}(10)$eV$^2$, and furthermore $\sim \Big{(}\frac{R}{3m_s^2}\Big{)}^{\delta-1}\sim
 \mathcal{O}(10^{-111})$eV$^2$ while $\sim \Big{(}\frac{R}{3m_s^2}\Big{)}^{\delta-2}\sim
 \mathcal{O}(10^{-223})$eV$^2$. Also since $V_0\ll R^2/M^2$, the potential term is also negligible compared
to the $R^2$ term. Also since primordially, the scalar field loses
its energy and transfers it to the dark matter fluid, as it can be
seen from Eqs. (\ref{fluidcontinuityequations} and
(\ref{interactionterm3}), the kinetic energy term for the scalar
field can also be neglected. In addition, the dark matter fluid
energy density redshifts as $\sim a^{-3}$ while the radiation
fluid redshifts as $\sim a^{-4}$ and since during inflation $a\sim
e^{\int_{t_i}^{t_f}H\mathrm{d}t}=e^N$, since the inflationary era
lasts for around $N\sim 60$ $e$-foldings, the matter and radiation
energy densities are also negligible in the field equations. Thus
only the $R^2$ gravity terms prevail and therefore, the field
equations become,
\begin{equation}\label{friedmanequationinflationaux}
3H^2\left(1+\frac{2}{M^2}R\right)=\frac{R^2}{2M}-\frac{6H\dot{R}}{M^2}\,
,
\end{equation}
or equivalently,
\begin{equation}\label{patsunappendixinflation}
3\ddot{H}-3\frac{\dot{H}^2}{H}+\frac{2M^2H}{6}=-9H\dot{H}\, ,
\end{equation}
which when solved yield an approximate quasi-de Sitter evolution,
\begin{equation}\label{quasidesitter}
H(t)=H_0-\frac{M^2}{36} t\, .
\end{equation}
The phenomenology of the Jordan frame vacuum $R^2$ model with the
quasi-de Sitter evolution produces a viable inflationary era,
compatible with the latest Planck data \cite{Planck:2018vyg},
since the spectral index as a function of the $e$-foldings number
is $n_s\sim 1-\frac{2}{N}$ and the predicted tensor-to-scalar
ratio is $r\sim \frac{12}{N^2}$. It is worth discussing how the
$F(R)$ gravity inflationary phenomenology is obtained, since we
will also need the exact value of the tensor spectral index in
order to calculate the energy spectrum of the primordial
gravitational waves. Assuming that the slow-roll conditions apply
during the inflationary era,
\begin{equation}\label{slowrollconditionshubble}
\ddot{H}\ll H\dot{H},\,\,\, \frac{\dot{H}}{H^2}\ll 1\, ,
\end{equation}
the dynamical evolution of inflation for a general $F(R)$ gravity
is quantified by the slow-roll indices $\epsilon_1$ ,$\epsilon_2$,
$\epsilon_3$, $\epsilon_4$, which are
\cite{Hwang:2005hb,reviews1,reviews6},
\begin{equation}
\label{restofparametersfr}\epsilon_1=-\frac{\dot{H}}{H^2}, \quad
\epsilon_2=0\, ,\quad \epsilon_3= \frac{\dot{F}_R}{2HF_R}\, ,\quad
\epsilon_4=\frac{\ddot{F}_R}{H\dot{F}_R}\,
 ,
\end{equation}
and the spectral index of the primordial scalar perturbations and
the tensor-to-scalar ratio are written as follows
\cite{reviews1,Hwang:2005hb},
\begin{equation}
\label{epsilonall} n_s=
1-\frac{4\epsilon_1-2\epsilon_3+2\epsilon_4}{1-\epsilon_1},\quad
r=48\frac{\epsilon_3^2}{(1+\epsilon_3)^2}\, .
\end{equation}
Using the Raychaudhuri equation for $F(R)$ gravity, we obtain,
\begin{equation}\label{approx1}
\epsilon_1=-\epsilon_3(1-\epsilon_4)\, ,
\end{equation}
therefore we have approximately,
\begin{equation}
\label{spectralfinal} n_s\simeq 1-6\epsilon_1-2\epsilon_4\, ,
\end{equation}
and also,
\begin{equation}
\label{tensorfinal} r\simeq 48\epsilon_1^2\, .
\end{equation}
Also considering $\epsilon_4=\frac{\ddot{F}_R}{H\dot{F}_R}$ we
get,
\begin{equation}\label{epsilon41}
\epsilon_4=\frac{\ddot{F}_R}{H\dot{F}_R}=\frac{\frac{d}{d
t}\left(F_{RR}\dot{R}\right)}{HF_{RR}\dot{R}}=\frac{F_{RRR}\dot{R}^2+F_{RR}\frac{d
(\dot{R})}{d t}}{HF_{RR}\dot{R}}\, ,
\end{equation}
and due to the fact that,
\begin{equation}\label{rdot}
\dot{R}=24\dot{H}H+6\ddot{H}\simeq 24H\dot{H}=-24H^3\epsilon_1\, ,
\end{equation}
combined with Eq. (\ref{epsilon41}) we get,
\begin{equation}\label{epsilon4final}
\epsilon_4\simeq -\frac{24
F_{RRR}H^2}{F_{RR}}\epsilon_1-3\epsilon_1+\frac{\dot{\epsilon}_1}{H\epsilon_1}\,
.
\end{equation}
By using,
\begin{equation}\label{epsilon1newfiles}
\dot{\epsilon}_1=-\frac{\ddot{H}H^2-2\dot{H}^2H}{H^4}=-\frac{\ddot{H}}{H^2}+\frac{2\dot{H}^2}{H^3}\simeq
2H \epsilon_1^2\, ,
\end{equation}
$\epsilon_4$ becomes,
\begin{equation}\label{finalapproxepsilon4}
\epsilon_4\simeq -\frac{24
F_{RRR}H^2}{F_{RR}}\epsilon_1-\epsilon_1\, .
\end{equation}
The tensor spectral index is equal to
\cite{Hwang:2005hb,reviews1,Odintsov:2020thl},
\begin{equation}\label{tensorspectralindexr2gravity}
n_{\mathcal{T}}\simeq -2 (\epsilon_1+\epsilon_3)\, ,
\end{equation}
hence by using Eq. (\ref{finalapproxepsilon4}) we get,
\begin{equation}\label{tensorspectralindexr2ini}
n_{\mathcal{T}}\simeq -2 \frac{\epsilon_1^2}{1+\epsilon_1}\simeq
-2\epsilon_1^2\, .
\end{equation}
For the case at hand, which is the $R^2$ model, we get,
\begin{equation}\label{r2modeltensorspectralindexfinal}
n_{\mathcal{T}}\simeq -\frac{1}{2N^2}\, ,
\end{equation}
therefore for $N\sim 60$ we get $n_{\mathcal{T}}=-0.000138889$,
$n_s\simeq 0.963$ and finally $r\simeq 0.0033$, which shall be
used in the analysis of the energy spectrum of the primordial
gravitational waves in the next section.

\subsection{Post-inflationary Evolution of the Universe and the Phase Space of the Scalar-dark matter Fluids Subsystem}

After the inflationary era and during reheating and thereafter,
the $F(R)$ gravity terms cease to dominate the evolution, and the
radiation fluid, the matter fluid and the scalar field fluid start
to control the evolution.
\begin{table}[h!]
  \begin{center}
    \caption{\emph{Fixed Points of the Dynamical System (\ref{dynamicalsystemtwodimensionaldynsubsystemain}) for General Values of $\beta$ and $\lambda$.}}
    \label{table1}
    \begin{tabular}{|r|r|}
     \hline
      \textbf{Name of Fixed Point} & \textbf{Fixed Point Values for General $\beta$ and $\lambda$}  \\
           \hline
      $P_1^*$ & $(x_*,y_*)=(-1,0)$  \\ \hline
      $P_2^*$ & $(x_*,y_*)=(1,0)$\\ \hline
      $P_3^*$ & $(x_*,y_*)=(\frac{2 \beta }{3},0)$ \\ \hline
      $P_4^*$ & $(x_*,y_*)=(\frac{\lambda }{\sqrt{6}},-\frac{\sqrt{2 \beta  \lambda ^2-12 \beta -\sqrt{6} \lambda ^3+6 \sqrt{6} \lambda }}{\sqrt{6} \sqrt{\sqrt{6} \lambda -2 \beta }})$
      \\\hline
      $P_5^*$ & $(x_*,y_*)=(\frac{\lambda }{\sqrt{6}},\frac{\sqrt{2 \beta  \lambda ^2-12 \beta -\sqrt{6} \lambda ^3+6 \sqrt{6} \lambda }}{\sqrt{6} \sqrt{\sqrt{6} \lambda -2 \beta }})$
      \\\hline
      $P_6^*$ & $(x_*,y_*)=(-\frac{3}{2 \beta -\sqrt{6} \lambda },-\frac{\sqrt{\frac{6 \lambda ^2}{2 \beta -\sqrt{6} \lambda }-\frac{2 \sqrt{6} \beta  \lambda }{2 \beta -\sqrt{6} \lambda }-\frac{18}{2 \beta -\sqrt{6} \lambda }-4 \beta +\sqrt{6} \lambda }}{\sqrt{2} \sqrt{\sqrt{6} \lambda -2 \beta }})$
      \\\hline
      $P_7^*$ & $(x_*,y_*)=(-\frac{3}{2 \beta -\sqrt{6} \lambda },\frac{\sqrt{\frac{6 \lambda ^2}{2 \beta -\sqrt{6} \lambda }-\frac{2 \sqrt{6} \beta  \lambda }{2 \beta -\sqrt{6} \lambda }-\frac{18}{2 \beta -\sqrt{6} \lambda }-4 \beta +\sqrt{6} \lambda }}{\sqrt{2} \sqrt{\sqrt{6} \lambda -2 \beta }})$
      \\\hline
    \end{tabular}
  \end{center}
\end{table}
In the way we chose the interaction between the matter and the
scalar field fluids, the scalar field fluid loses its energy
primordially which transfers it to the matter fluid, thus the
radiation and the dark matter fluid control the evolution up to
the point that the interaction between the scalar and dark matter
fluid flips its sign, see Eq. (\ref{interactionterm1}). This sign
flip occurs during the radiation domination era, and well before
the matter-radiation equality, and note that at the critical
matter density $\rho_m^c$, the interaction switches off.
Apparently, the two-dimensional system composed by the scalar
field and matter fluids controls the evolution after the critical
matter density $\rho_m^c$, competing with the radiation fluid.
Thus in this section we shall analyze the two-dimensional scalar
field-matter fluid phase space in order to reveal the possible
dynamical evolution of the Universe after the critical density
$\rho_m^c$. As we show, essentially, there might be deformations
in the total EoS parameter during the radiation domination era,
caused by the matter-scalar field interactions.
\begin{table}[h!]
  \begin{center}
    \caption{\emph{Fixed Points of the Dynamical System (\ref{dynamicalsystemtwodimensionaldynsubsystemain}) for $\beta=2$ and $\lambda=\sqrt{3}$.}}
    \label{table2}
    \begin{tabular}{|r|r|}
     \hline
      \textbf{Name of Fixed Point} & \textbf{Fixed Point Values for $\beta=2$ and $\lambda=\sqrt{3}$}  \\
           \hline
      $P_1^*$ & $(x_*,y_*)=(-1,0)$  \\\hline
      $P_2^*$ & $(x_*,y_*)=(1,0)$\\\hline
      $P_3^*$ & $(x_*,y_*)=(\frac{4}{3},0)$ \\\hline
      $P_4^*$ & $(x_*,y_*)=(\frac{1}{\sqrt{2}},-\frac{1}{\sqrt{2}})$ \\\hline
      $P_5^*$ & $(x_*,y_*)=(\frac{1}{\sqrt{2}},\frac{1}{\sqrt{2}})$ \\\hline
      $P_6^*$ & $(x_*,y_*)=(6+\frac{9}{\sqrt{2}},-\left(2+\frac{3}{\sqrt{2}}\right) \sqrt{25-12 \sqrt{2}})$ \\\hline
      $P_7^*$ & $(x_*,y_*)=(6+\frac{9}{\sqrt{2}},\left(2+\frac{3}{\sqrt{2}}\right) \sqrt{25-12 \sqrt{2}})$ \\\hline
    \end{tabular}
  \end{center}
\end{table}
As we already mentioned, we shall focus on quintessence type
potentials for the scalar field, of the form given in Eq.
(\ref{exponentialpotential}). We shall study the matter
fluid-scalar field fluid two-dimensional phase space and its
dynamics. A crucial assumption for our analysis is that at the
moment when the interaction between the scalar field and the dark
matter fluid is switched off, thus at the critical matter density
$\rho_m^c$, the scalar field has a constant EoS parameter and
satisfies,
\begin{equation}\label{eoscondition}
\dot{\phi}^2=\beta V(\phi)\, ,
\end{equation}
thus,
\begin{equation}\label{dddotphi}
\ddot{\phi}=\frac{\beta V'}{2}\, ,
\end{equation}
therefore, the field equation of the scalar field with no
interaction between the scalar and matter fluid,
\begin{equation}\label{fieldfree}
\ddot{\phi}+3H\dot{\phi}+V'=0\, ,
\end{equation}
yields,
\begin{equation}\label{eqnfreescalarkin}
\left(\frac{\beta+2}{2} \right)^2\left(
V'\right)^2=9H^2\dot{\phi}^2\, ,
\end{equation}
which yields,
\begin{equation}\label{potentialapprox}
V=V_0e^{-\sqrt{\frac{6\beta}{\beta+2}}\kappa \phi}\, .
\end{equation}
Thus by comparing Eqs. (\ref{exponentialpotential}) and
(\ref{potentialapprox}) we obtain,
\begin{equation}\label{labdaconstraintbeta}
\lambda=\sqrt{\frac{6\beta}{\beta+2}}\, .
\end{equation}
Now the EoS parameter of the scalar field is defined to be,
\begin{equation}\label{eosscalardef}
w_{\phi}=\frac{\frac{\dot{\phi}^2}{2}-V}{\frac{\dot{\phi}^2}{2}+V}\,
,
\end{equation}
thus at the critical matter density $\rho_m^c$ where Eq.
(\ref{eoscondition}) holds true, the scalar field EoS becomes,
\begin{equation}\label{eosscalarfinal}
w_{\phi}=\frac{\beta-2}{\beta+2}\, .
\end{equation}
The crucial assumption we shall make is that the EoS parameter for
the scalar field at critical matter density $\rho_m^c$ is equal to
zero, that is,
\begin{equation}\label{assumptionmatterradiationequality}
w_{\phi}=0\,,\,\,\,\mathrm{at}\,\,\,\mathrm{matter}-\mathrm{radiation}\,\,\,\mathrm{equality}
\, .
\end{equation}
Thus in view of Eq. (\ref{eosscalarfinal}), we get $\beta=2$, and
also due to Eq. (\ref{labdaconstraintbeta}) we get that
$\lambda=\sqrt{3}$, that is,
\begin{equation}\label{valuesbetalambda}
\beta=2,\,\,\,\lambda=\sqrt{3}\, .
\end{equation}
This is crucial for the dynamical system analysis that follows.
\begin{table}[h!]
  \begin{center}
    \caption{\emph{Eigenvalues of the Jacobian matrix for the dynamical system (\ref{dynamicalsystemtwodimensionaldynsubsystemain}) for $\beta=2$ and $\lambda=\sqrt{3}$.}}
    \label{table3}
    \begin{tabular}{|r|r|r|}
     \hline
      \textbf{Name of Fixed Point} & \textbf{Eigenvalues} & \textbf{Stability}  \\
           \hline
      $P_1^*$ & $(-1,\frac{1}{2} (-3) \left(\sqrt{2}-2\right))$ & unstable
      \\\hline
      $P_2^*$ & $(7,\frac{3}{2} \left(\sqrt{2}+2\right))$ &
      unstable\\\hline
      $P_3^*$ & $(\frac{1}{6} \left(25-12 \sqrt{2}\right),\frac{7}{6})$  &
      unstable\\\hline
      $P_4^*$ & $-2 \sqrt{2},-\frac{3}{2})$ & stable\\ \hline
      $P_5^*$ & $(-2 \sqrt{2},-\frac{3}{2})$ & stable \\ \hline
      $P_6^*$ & $(47.5601, -24.3322)$ & unstable \\ \hline
      $P_7^*$ & $(47.5601, -24.3322)$ & unstable\\ \hline
    \end{tabular}
  \end{center}
\end{table}
Now let us construct the autonomous dynamical system for the
scalar field-matter fluid two dimensional subsystem that controls
the dynamics near the critical matter density $\rho_m^c$ and
thereafter. In the literature such scalar field-fluid systems have
been studied in the literature with \cite{Boehmer:2008av} and
without interaction \cite{Copeland:1997et}. The Friedmann equation
for the scalar-matter fluid two dimensional subsystem that
dominates the evolution is,
\begin{equation}\label{subsystemfriedmann}
3H^2=\kappa^2\rho_m+\frac{\kappa^2\dot{\phi}^2}{2}+V\, ,
\end{equation}
which can be rewritten as,
\begin{equation}\label{friedmannconstraint}
\Omega_m+\Omega_{\phi}=1\, ,
\end{equation}
where,
\begin{equation}\label{omegaphiomegam}
\Omega_{\phi}=\frac{\kappa^2\rho_{\phi}}{3H^2},\,\,\,\Omega_m=\frac{\kappa^2\rho_{m}}{3H^2}\,
.
\end{equation}
The total EoS parameter $w_{tot}$ is equal to,
\begin{equation}\label{totaleosparameter}
w_{tot}=\frac{P_{\phi}}{\rho_{\phi}+\rho_m}=w_{\phi}\Omega_{\phi}\,
,
\end{equation}
and the total energy satisfies the continuity equation,
\begin{equation}\label{continuitytotal}
\dot{\rho}_{tot}+3H(1+w_{tot})\rho_{tot}=0\, ,
\end{equation}
while the interacting scalar-dark matter fluids have the following
continuity equations,
\begin{align}\label{fluidcontinuityequations1}
& \dot{\rho}_m+3H \rho_m=-Q\, , \\ \notag &
\dot{\rho}_{\phi}+3H(\rho_{\phi}+P_{\phi})=Q\, ,
\end{align}
with $Q$ being defined in Eq. (\ref{interactionterm1}). We
introduce the dimensionless variables,
\begin{equation}\label{variablesdynamicalsystem}
x=\frac{\kappa^2\dot{\phi}^2}{6H^2},\,\,\,y=\frac{\kappa^2V}{3H^2}\,
,
\end{equation}
and using these we have,
\begin{equation}\label{constraintsnew}
w_{\phi}=\frac{x^2-y^2}{x^2+y^2},\,\,\,\Omega_{\phi}=x^2+y^2\leq
1\, ,
\end{equation}
while the Raychaudhuri equation is written,
\begin{equation}\label{ryachaudhuru}
-2\dot{H}=3H^2(1+x^2-y^2)\, .
\end{equation}
Using the field equations, the continuity equations
(\ref{fluidcontinuityequations1}) and the variables
(\ref{variablesdynamicalsystem}), we can form the following
two-dimensional fully autonomous dynamical system
\cite{Boehmer:2008av},
\begin{align}\label{dynamicalsystemtwodimensionaldynsubsystemain}
& \frac{\mathrm{d}x}{\mathrm{d}N}=-3x+\frac{\lambda \sqrt{6}}{2}y^2+\frac{3 x}{2}\left(1+x^2-y^2 \right)+\beta \left(1-x^2-y^2\right)\, ,              \\
\notag & \frac{\mathrm{d}x}{\mathrm{d}N}=-\frac{\lambda
\sqrt{6}}{2}x\,y+\frac{3 y}{2}\left(1+x^2-y^2 \right)\, ,
\end{align}
where instead of the cosmic time we used the $e$-foldings number
$N$ as a dynamical variable. Recall that in our case, the
parameters $\beta$ and $\lambda$ are fixed by our assumptions to
have the values appearing in Eq. (\ref{valuesbetalambda}). The
dynamical system
(\ref{dynamicalsystemtwodimensionaldynsubsystemain}) is autonomous
and can easily be studied. First let us present the fixed points
of this dynamical system expressed in terms of general values of
$\beta$ and $\lambda$ and then we specify for the values appearing
in Eq. (\ref{valuesbetalambda}). The fixed points of the dynamical
system (\ref{dynamicalsystemtwodimensionaldynsubsystemain}) for
general $\lambda$ and $\beta$ are given in Table \ref{table1},
while for the values of $\beta$ and $\lambda$ specified in Eq.
(\ref{valuesbetalambda}), the fixed points are given in Table
\ref{table2}.

\begin{figure}[h!]
\centering
\includegraphics[width=20pc]{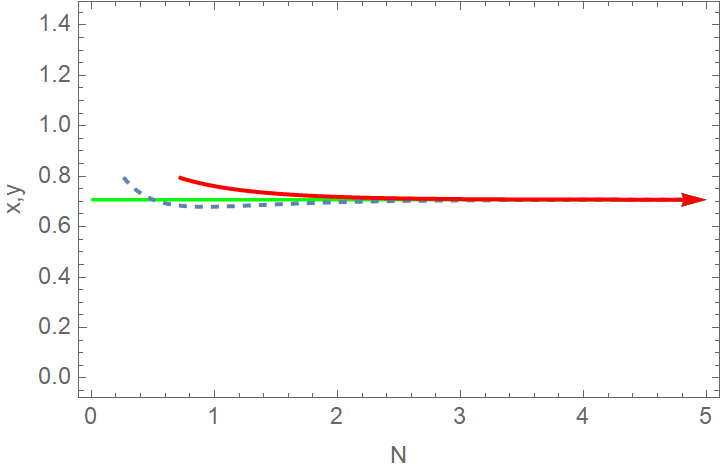}
\includegraphics[width=20pc]{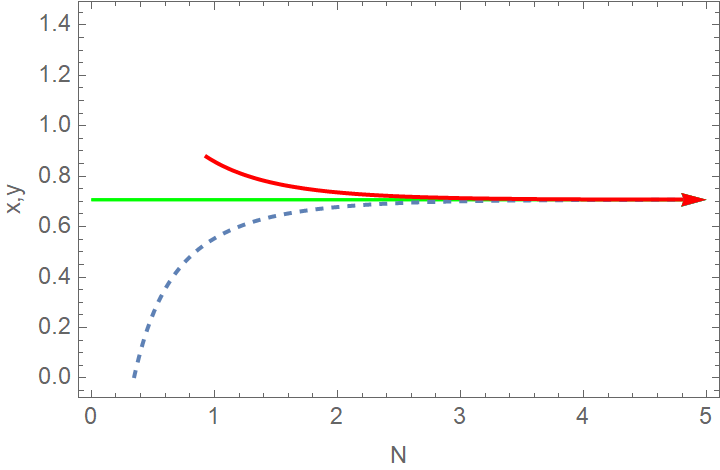}
\includegraphics[width=20pc]{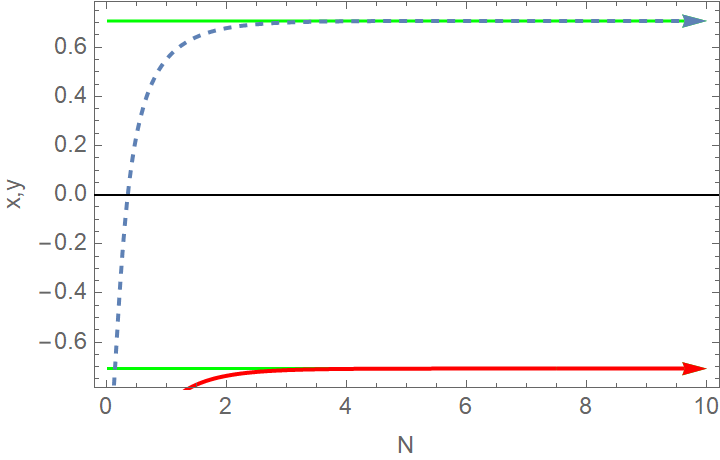}
\includegraphics[width=20pc]{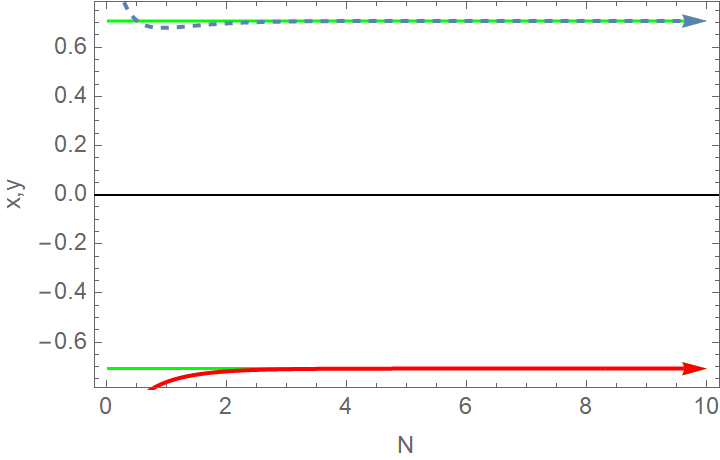}
\caption{Trajectories $x(N)$ (blue dashed curve) and $y(N)$ (red
thick curve) in the phase space of the dynamical system
(\ref{dynamicalsystemtwodimensionaldynsubsystemain}) for various
initial conditions. The green lines indicate the values
$1/\sqrt{2}$ and $-1/\sqrt{2}$. As it can be seen, the stable dark
matter attractors $P_4^*$ and $P_5^*$ are reached quite fast in
the phase space.} \label{plot1}
\end{figure}

\begin{figure}[h!]
\centering
\includegraphics[width=25pc]{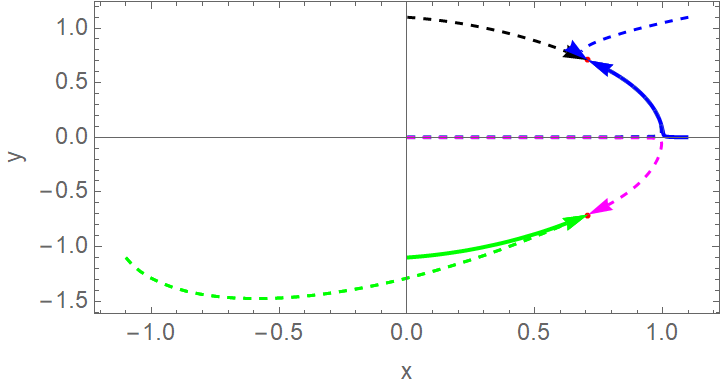}
\caption{Trajectories in the $x(N)-y(N)$ plane of the phase space
of the dynamical system
(\ref{dynamicalsystemtwodimensionaldynsubsystemain}) for various
initial conditions. Note the magenta dashed and the blue thick
curves, which both pass through the unstable fixed point $P_2^*$
before ending to the stable dark matter attractors $P_4^*$ and
$P_5^*$ respectively.} \label{plot2}
\end{figure}

Now let us address the stability of the fixed points appearing in
Table \ref{table2}. The eigenvalues of the Jacobian matrix are
given in Table \ref{table3}. As it can be seen, only the fixed
points $P_4^*$ and $P_5^*$ are stable, but let us comment that not
all the fixed points are physically acceptable satisfying the
Friedmann constraint. In Table \ref{table4} we gather the values
of the physical parameters at the fixed points, and as it can be
seen, the fixed points $P_3^*$, $P_6^*$ and $P_7^*$ are
unphysical. So let us focus on the four other physical points,
which describe interesting physical evolution dynamics.
Specifically, fixed points $P_4^*$ and $P_5^*$ describe stable
dark matter dominated attractors while the fixed points $P_1^*$
and $P_2^*$ describe unstable kination attractors, as it can be
seen in Table \ref{table4}. The fixed points $P_4^*$ and $P_5^*$
are not identical, but describe the same dark matter dominated
physics, and the same applies for the fixed points $P_1^*$ and
$P_2^*$ which describe the same kination domination physics. Thus
the phase space of the dynamical system
(\ref{dynamicalsystemtwodimensionaldynsubsystemain}) is deemed
quite intriguing from a physical point of view. As it seems, the
scalar field takes the energy from the matter perfect fluid, and
can lead the dynamical system eventually to stable dark matter
attractors generated by the scalar field itself. But more
remarkable and of profound physical importance is that the
dynamical system may be attracted to kination dominated fixed
points, which due to the fact that these are unstable, the
dynamical system is eventually repelled from the kination fixed
points and finally ends up to the stable dark matter attractors.
Thus the dynamical system eventually is described by a matter
dominated era controlled by the scalar field, during the radiation
domination era, so the total EoS of the radiation domination era
is disturbed and thus can be larger than $w=1/3$ and closer to the
stiff evolution value $w=1$. This said behavior may continue until
matter dominates and after that, the $F(R)$ gravity terms start to
dominate the late-time dynamics and generate the dark energy era.
Hence, there is an obvious probability that there might exist a
set of initial conditions in the Universe, that may lead to a
scalar field kination eras, and thus deformations of the radiation
domination era, well before the BBN era. This probability must be
examined numerically by solving the dynamical system using various
sets of initial conditions, and if our predictions are correct,
before the final dark matter attractors are reached, the dynamical
system composed by the scalar field and the dark matter fluid may
pass from the kinetic dominated fixed points. Let us first show
numerically that the dynamical system ends up to the stable dark
matter attractors. We solve the dynamical system
(\ref{dynamicalsystemtwodimensionaldynsubsystemain}) numerically
for various initial conditions and we present the behavior of the
trajectories $(x(N),y(N))$ in the phase space as a function of the
$e$-foldings in Fig. \ref{plot1}. The blue dashed curves represent
$x(N)$ while the red thick curves represent the trajectories
$y(N)$. The green lines indicate the values $1/\sqrt{2}$ and
$-1/\sqrt{2}$. As it can be seen, the stable dark matter
attractors $P_4^*$ and $P_5^*$ are reached quite fast in the phase
space.
\begin{table}[h!]
  \begin{center}
    \caption{\emph{Values of the physical parameters for the fixed points of the  dynamical system (\ref{dynamicalsystemtwodimensionaldynsubsystemain}) for $\beta=2$ and $\lambda=\sqrt{3}$.}}
    \label{table4}
    \begin{tabular}{|r|r|r|r|r|r|}
     \hline
      \textbf{Name of Fixed Point} & $w_{tot}$ & $\Omega_{\phi}$ & $w_{\phi}$ & $\Omega_m$ &  \textbf{Stability}
      \\ \hline
      $P_1^*$ & 1 & 1 & 1 & 0 & unstable \\ \hline
      $P_2^*$ & 1 & 1 & 1 & 0 & unstable\\ \hline
      $P_3^*$ & 1.77778  & 1.77 & 1 & -0.33 & unstable\\ \hline
      $P_4^*$ & 0 & 1 & 0 & 0 & stable\\ \hline
      $P_5^*$ & 0 & 1 & 0 & 0 & stable \\ \hline
      $P_6^*$ & 16.4853 & 289.25 & 0.0569932 & -288.25 & unstable \\\hline
      $P_7^*$ & 16.4853 & 289.25 & 0.0569932 & -288.25 & unstable\\
      \hline
    \end{tabular}
  \end{center}
\end{table}
But the plots in Fig. \ref{plot1} do not allow us to see
explicitly whether the kinetic dominated fixed points $P_1^*$ and
$P_2^*$ are reached in the phase, thus we will use a parametric
plot in the plane $x(N)-y(N)$ to see whether there exist initial
conditions in the phase space that generate trajectories which
pass through the kination fixed points $P_1^*$ and $P_2^*$ before
they end up to the stable dark matter attractors $P_4^*$ and
$P_5^*$.
In Fig. \ref{plot2} we present the trajectories of the dynamical
system (\ref{dynamicalsystemtwodimensionaldynsubsystemain}) in the
$x(N)-y(N)$ plane for various initial conditions. As it can be
seen, there exist various trajectories in the phase space but the
most interesting for our scenario are the magenta dashed one and
the blue thick curves, which both pass through the unstable fixed
point $P_2^*$ before ending to the stable dark matter attractors
$P_4^*$ and $P_5^*$ respectively.
Apparently, our theoretical prediction that the dynamical system
composed by the matter and scalar field fluids may experience a
short stiff evolution after the critical matter density
$\rho_m^c$, during the radiation domination era, before the BBN,
is a probably realistic scenario, which may have profound
observational implications, regarding gravitational wave physics.
This is the subject of the next section.

Let us recapitulate at this point our findings. We initially
assumed a non-trivial interaction between the matter-scalar field
fluids, which after critical matter density $\rho_m^c$, reached
during the radiation domination era by the dark matter fluid,
makes the scalar field fluid to gain energy from the dark matter
fluid. Since these two fluids dominate the evolution during the
radiation domination era, we studied the two-dimensional phase
space formed by these two fluids. We demonstrated that there exist
two stable dark matter dominated fixed points, and two unstable
kination dominated fixed points, all realized by the scalar field.
We analyzed the trajectories in the phase space and we showed that
there exist trajectories which pass from the unstable kination
fixed point $P_2^*$ before they end up to the dark matter fixed
points. Thus it is possible that the total EoS of the Universe
during the radiation domination era might be deformed and can
actually be larger than $w=1/3$. This scalar field originating EoS
deformations of the radiation domination era may have profound
observational implications related to the energy spectrum of the
primordial gravitational waves. In the next section we shall
analyze this possibility in some detail.

\section{Radiation Domination EoS Deformations and the Energy Spectrum of the Primordial Gravitational Waves}

After the poor findings in the Large Hadron Collider in the
post-Higgs discovery, the focus of theoretical physicists has
turned to the sky and specifically to CMB related and
gravitational waves related experiments. There is a large stream
of articles in the literature on primordial gravitational waves,
see for example Refs.
\cite{Kamionkowski:2015yta,Turner:1993vb,Boyle:2005se,Zhang:2005nw,Caprini:2018mtu,Clarke:2020bil,Smith:2005mm,Giovannini:2008tm,Liu:2015psa,Vagnozzi:2020gtf,Giovannini:2023itq,Giovannini:2022eue,Giovannini:2022vha,Giovannini:2020wrx,Giovannini:2019oii,Giovannini:2019ioo,Giovannini:2014vya,Giovannini:2009kg,Kamionkowski:1993fg,Giare:2020vss,Zhao:2006mm,Lasky:2015lej,
Cai:2021uup,Odintsov:2021kup,Lin:2021vwc,Zhang:2021vak,Visinelli:2017bny,Pritchard:2004qp,Khoze:2022nyt,Casalino:2018tcd,Oikonomou:2022xoq,Casalino:2018wnc,ElBourakadi:2022anr,Sturani:2021ucg,Vagnozzi:2022qmc,Arapoglu:2022vbf,Giare:2022wxq,Oikonomou:2021kql,Gerbino:2016sgw,Breitbach:2018ddu,Pi:2019ihn,Khlopov:2023mpo}
and the recent review Ref. \cite{Odintsov:2022cbm} and references
therein. Regarding the energy spectrum of the gravitational waves,
taking into account a  standard radiation domination era followed
by a dark matter era, and the latter followed by a dark energy
era, this is equal to,
\begin{equation}
    \Omega_{\rm gw}(f)= \frac{k^2}{12H_0^2}\Delta_h^2(k),
    \label{GWspec}
\end{equation}
with $\Delta_h^2(k)$ being equal to,
\begin{equation}
\Delta_h^2(k)=\Delta_h^{({\rm p})}(k)^{2} \left (
\frac{\Omega_m}{\Omega_\Lambda} \right )^2 \left (
\frac{g_*(T_{\rm in})}{g_{*0}} \right ) \left (
\frac{g_{*s0}}{g_{*s}(T_{\rm in})} \right )^{4/3} \left
(\overline{ \frac{3j_1(k\tau_0)}{k\tau_0} } \right )^2 T_1^2\left
( x_{\rm eq} \right ) T_2^2\left ( x_R \right )\, ,
\label{mainfunctionforgravityenergyspectrum}
\end{equation}
and the oscillating term must be calculated for a Hubble time, and
in addition $\Delta_h^{({\rm p})}(k)^{2}$ stands for the
primordial tensor power spectrum of the inflationary era, and it
is equal to,
\begin{equation}
\Delta_h^{({\rm
p})}(k)^{2}=\mathcal{A}_T(k_{ref})\left(\frac{k}{k_{ref}}
\right)^{n_{\mathcal{T}}}\, .
\label{primordialtensorpowerspectrum}
\end{equation}
The above must be calculated at the CMB pivot scale which we
assume it is $k_{ref}=0.002$$\,$Mpc$^{-1}$, and $n_{\mathcal{T}}$
denotes the tensor spectral index, while $\mathcal{A}_T(k_{ref})$
stands for amplitude of the tensor perturbations amplitude which
can be expressed in terms of the amplitude of the scalar
perturbations $\mathcal{P}_{\zeta}(k_{ref})$ in the following way,
\begin{equation}\label{amplitudeoftensorperturbations}
\mathcal{A}_T(k_{ref})=r\mathcal{P}_{\zeta}(k_{ref})\, ,
\end{equation}
with $r$ being the tensor-to-scalar ratio. Hence,
\begin{equation}\label{primordialtensorspectrum}
\Delta_h^{({\rm
p})}(k)^{2}=r\mathcal{P}_{\zeta}(k_{ref})\left(\frac{k}{k_{ref}}
\right)^{n_{\mathcal{T}}}\, .
\end{equation}
Note that the transfer function $T_1(x_{\rm eq})$ in Eq.
(\ref{mainfunctionforgravityenergyspectrum}) directly connects the
energy spectrum at present day  with the modes $k$ which reentered
the Hubble horizon during the matter-radiation equality, and this
is equal to,
\begin{equation}
    T_1^2(x_{\rm eq})=
    \left [1+1.57x_{\rm eq} + 3.42x_{\rm eq}^2 \right ], \label{T1}
\end{equation}
where $x_{\rm eq}=k/k_{\rm eq}$ and $k_{\rm eq}\equiv a(t_{\rm
eq})H(t_{\rm eq}) = 7.1\times 10^{-2} \Omega_m h^2$ Mpc$^{-1}$.
Furthermore, the other transfer function $T_2(x_R)$ directly
connects the energy spectrum of the gravitational waves at present
day to the one corresponding to the era that the mode $k$
reentered the Hubble horizon during the reheating era and before
it ended, therefore when $k>k_R$, and the transfer function is
equal to,
\begin{equation}\label{transfer2}
 T_2^2\left ( x_R \right )=\left(1-0.22x^{1.5}+0.65x^2
 \right)^{-1}\, ,
\end{equation}
with $x_R=\frac{k}{k_R}$, while the reheating temperature
wavenumber $k_R$ is equal to,
\begin{equation}
    k_R\simeq 1.7\times 10^{13}~{\rm Mpc^{-1}}
    \left ( \frac{g_{*s}(T_R)}{106.75} \right )^{1/6}
    \left ( \frac{T_R}{10^6~{\rm GeV}} \right )\, ,  \label{k_R}
\end{equation}
where $T_R$ stands for the reheating temperature. Note that for
the energy spectrum of the gravitational waves at present day we
took into account the overall damping effect in the early Universe
generated by the non-constancy of the total number of the
relativistic degrees of freedom in which case the scale factor
behaves as $a(t) \propto T^{-1}$ \cite{Watanabe:2006qe}, hence the
total damping factor due to this behaves as,
\begin{equation}
    \left ( \frac{g_*(T_{\rm in})}{g_{*0}} \right )
    \left ( \frac{g_{*s0}}{g_{*s}(T_{\rm in})} \right )^{4/3},
\end{equation}
with $T_{\rm in}$ denotes the temperature at the horizon reentry,
\begin{equation}
    T_{\rm in}\simeq 5.8\times 10^6~{\rm GeV}
    \left ( \frac{g_{*s}(T_{\rm in})}{106.75} \right )^{-1/6}
    \left ( \frac{k}{10^{14}~{\rm Mpc^{-1}}} \right ).
\end{equation}
Note that the reheating temperature is basically an unknown free
parameter  in the above context. Furthermore,
$g_*(T_{\mathrm{in}}(k))$ in Eq.
(\ref{mainfunctionforgravityenergyspectrum}) is equal to
\cite{Kuroyanagi:2014nba},
\begin{align}\label{gstartin}
& g_*(T_{\mathrm{in}}(k))=g_{*0}\left(\frac{A+\tanh \left[-2.5
\log_{10}\left(\frac{k/2\pi}{2.5\times 10^{-12}\mathrm{Hz}}
\right) \right]}{A+1} \right) \left(\frac{B+\tanh \left[-2
\log_{10}\left(\frac{k/2\pi}{6\times 10^{-19}\mathrm{Hz}} \right)
\right]}{B+1} \right)\, ,
\end{align}
where $A$ and $B$ are equal to,
\begin{equation}\label{alphacap}
A=\frac{-1-10.75/g_{*0}}{-1+10.75g_{*0}}\, ,
\end{equation}
\begin{equation}\label{betacap}
B=\frac{-1-g_{max}/10.75}{-1+g_{max}/10.75}\, ,
\end{equation}
with $g_{max}=106.75$ and $g_{*0}=3.36$. Moreover
$g_{*0}(T_{\mathrm{in}}(k))$ can be calculated by simply replacing
$g_{*0}=3.36$ with $g_{*s}=3.91$ in Eqs. (\ref{gstartin}),
(\ref{alphacap}) and (\ref{betacap}). As a final comment, for the
calculation of the energy spectrum of the primordial gravitational
waves, we also took into account the damping factor $\sim
(\Omega_m/\Omega_\Lambda)^2$ due to the present day acceleration
of the Universe.

In the previous section we showed that it is possible for the
Universe to experience short deformations of its total EoS
parameter during the radiation domination era (can also be during
the reheating era). Thus let us see how these pre-BBN deformations
can affect the energy spectrum of the primordial gravitational
waves. We will assume that the EoS deformations occur when
wavenumbers of the order $k_s\sim 10^{11}\,$Mpc$^{-1}$ reenter the
Hubble horizon, so this era corresponds to the reheating era, or
at some point during the early radiation domination era. Since the
scalar field stiff attractors, affect the total EoS during the
reheating, we will assume that the total background EoS parameter
during this era is $w=0.8$, or even larger, somewhere in the range
$w\sim[1/3,1]$. The change of the background EoS parameter for a
dark matter dominated one, to the value $w$, has its imprint on
the energy spectrum of the gravitational waves, since an overall
multiplication factor of the form $\sim
\left(\frac{k}{k_{s}}\right)^{r_s}$ is included in the energy
spectrum, where $r_s=-2\left(\frac{1-3 w}{1+3 w}\right)$
\cite{Gouttenoire:2021jhk}. Therefore, the final expression for
the total $h^2$-scaled energy spectrum of the primordial
gravitational waves finally takes the form,
\begin{align}
\label{GWspecfRnewaxiondecay}
    &h^2\Omega_{\rm gw}(f)=S_k(f)\times \frac{k^2}{12H_0^2}r\mathcal{P}_{\zeta}(k_{ref})\left(\frac{k}{k_{ref}}
\right)^{n_{\mathcal{T}}} \left ( \frac{\Omega_m}{\Omega_\Lambda}
\right )^2
    \left ( \frac{g_*(T_{\rm in})}{g_{*0}} \right )
    \left ( \frac{g_{*s0}}{g_{*s}(T_{\rm in})} \right )^{4/3} \nonumber  \left (\overline{ \frac{3j_1(k\tau_0)}{k\tau_0} } \right )^2
    T_1^2\left ( x_{\rm eq} \right )
    T_2^2\left ( x_R \right )\, ,
\end{align}
where $S_k(f)$,
\begin{equation}\label{multiplicationfactor1}
S_k(f)=\left(\frac{k}{k_{s}}\right)^{r_s}\, ,
\end{equation}
and recall that $k_{ref}$ is the CMB pivot scale
$k_{ref}=0.002$$\,$Mpc$^{-1}$ and $n_{\mathcal{T}}$ stands for the
tensor spectral index of the primordial tensor perturbations,
while $r$ denotes the tensor-to-scalar ratio. It is important to
note once more that the reheating temperature is a free variable
and as it will prove once more, it will play an important role in
the final form of the predicted energy spectrum of the primordial
gravitational waves. Having all the necessary information at hand,
we now proceed to the determination of the predicted energy
spectrum of the primordial gravitational waves.
\begin{figure}[h!]
\centering
\includegraphics[width=40pc]{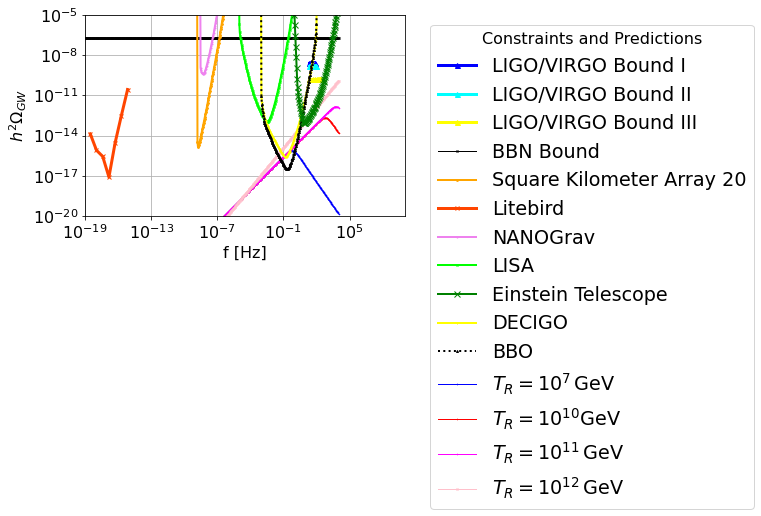}
\caption{The $h^2$-scaled gravitational wave energy spectrum for
the $R^2$ gravity driven inflationary era and a EoS deformations
of the reheating era which make the total EoS parameter take
values $w_{tot}>1/3$. We took $w_{tot}=0.8$ which occurs when the
modes with wavenumber $k=10^{11}\,$Mpc$^{-1}$ reenter the Hubble
horizon. The blue curve corresponds to the reheating temperature
$T_R=10^{7}\,$GeV, the red curve to $T_R=10^{10}\,$GeV, the
magenta curve to $T_R=10^{11}\,$GeV and the pink curve to
$T_R=10^{12}\,$GeV.} \label{plot3}
\end{figure}
In Fig. \ref{plot3} we present the $h^2$-scaled gravitational wave
energy spectrum for the model under study which contains a
primordial $R^2$ gravity driving the inflationary era and the
effects of EoS deformations with $w_{tot}=0.8$, of the reheating
(or early radiation domination) era which occurs when the modes
with wavenumber $k=10^{11}\,$Mpc$^{-1}$ reenter the Hubble
horizon, for four reheating temperatures, $T_R=10^{7}\,$GeV (blue
curve), $T_R=10^{10}\,$GeV (red curve), $T_R=10^{11}\,$GeV
(magenta curve) and $T_R=10^{12}\,$GeV (pink curve). For all the
plots we used the tensor spectral index and tensor-to-scalar ratio
of the $R^2$ model which we derived in the previous section. As it
can be seen, all the curves can be detected by the future DECIGO,
BBO, but only the large reheating temperature curve can be
detected by the Einstein telescope. This result is deemed quite
important since the energy spectrum of the pure $R^2$ gravity
cannot be detected by any of the future gravitational wave
experiments. Thus we demonstrated that short EoS deformations
occurring during the reheating era, can actually yield a
detectable gravitational wave energy spectrum.

\section*{Concluding Remarks and Discussion}

In this work we proposed a theoretical scenario in which the
Universe may pass through a brief reheating era deformations well
before the BBN era and well beyond the matter-radiation equality.
We used a model which is composed by an $F(R)$ gravity, the
radiation perfect fluid and an interacting system of dark matter
and scalar field fluids. The model is constructed in such a way so
that primordially and at late times the $F(R)$ gravity dominates
the evolution, thus producing the inflationary era and the dark
energy era, while in between, the Universe is dominated initially
by the radiation fluid and eventually after a critical matter
density $\rho_m^c$ during the reheating era or early radiation
domination era, by the interacting dark matter and scalar field
fluids which dominate over the radiation fluid, or cause
disturbance in its dominance over the evolution of the Universe.
The interaction between the dark matter and scalar fluids acts in
such a way so that after inflation, the scalar field fluid loses
its energy and transfers it to the dark matter fluid, and at the
critical matter density $\rho_m^c$ the interaction is switched of
effectively, while after the critical matter density $\rho_m^c$,
the interaction flips its sign and the scalar field gains energy
from the dark matter fluid. We formed the two-dimensional subspace
of the total cosmological phase space, composed by the dark matter
and scalar field fluids, and we constructed the autonomous
dynamical system that governs this phase space. We calculated the
fixed points and as we showed, there exist two unstable stiff
fixed points and two stable dark matter attractors. As we showed
numerically, there exist initial conditions for which the
trajectories in the phase space may pass through one of the two
kination fixed points, before ending up to the stable dark matter
attractors. This behavior makes possible for the Universe to
experience short deformations of its total EoS parameter, and we
examined the effects of such a short eras on the energy spectrum
of the primordial gravitational waves. As we showed, even with a
standard $R^2$ inflationary era, the predicted energy spectrum can
be detected by the future DECIGO and BBO experiments and in some
cases by the Einstein Telescope.

\section*{Acknowledgments}

This research has been is funded by the Committee of Science of
the Ministry of Education and Science of the Republic of
Kazakhstan (Grant No. AP19674478).


\begin{thebibliography}{99}


\bibitem{inflation1}
 A.~D.~Linde,
 %``Inflationary Cosmology,''
 Lect.\ Notes Phys.\ {\bf 738} (2008) 1
 %%%doi:10.1007/978-3-540-74353-8_1
 [arXiv:0705.0164 [hep-th]].
 %%CITATION = %%doi:10.1007/978-3-540-74353-8_1;%%
 %336 citations counted in INSPIRE as of 24 Dec 2016

\bibitem{inflation2} D.~S.~Gorbunov and V.~A.~Rubakov,
``Introduction to the theory of the early universe: Cosmological
perturbations and inflationary theory,'' Hackensack, USA: World
Scientific (2011) 489 p;
%6 citations counted in INSPIRE as of 14 Feb 2015
%
%V. Mukhanov, Physical foundations of cosmology, Cambridge, UK: Univ. Pr. (2005) 421 p; D. S. Gorbunov, V. A. Rubakov, Introduction to the theory of the early Universe: Cosmological perturbations and inflationary theory, Hackensack, USA, World Scientific (2011) 489 p


\bibitem{inflation3}A.~Linde,
%``Inflationary Cosmology after Planck 2013,''
arXiv:1402.0526 [hep-th];
%\cite{Lyth:1998xn}
%\bibitem{Lyth:1998xn}


\bibitem{inflation4}D.~H.~Lyth and A.~Riotto,
%``Particle physics models of inflation and the cosmological density perturbation,''
Phys.\ Rept.\  {\bf 314} (1999) 1 [hep-ph/9807278].








%\cite{CMB-S4:2016ple}
\bibitem{CMB-S4:2016ple}
K.~N.~Abazajian \textit{et al.} [CMB-S4],
%``CMB-S4 Science Book, First Edition,''
[arXiv:1610.02743 [astro-ph.CO]].
%1048 citations counted in INSPIRE as of 26 Oct 2021



%\cite{SimonsObservatory:2019qwx}
\bibitem{SimonsObservatory:2019qwx}
M.~H.~Abitbol \textit{et al.} [Simons Observatory],
%``The Simons Observatory: Astro2020 Decadal Project Whitepaper,''
Bull. Am. Astron. Soc. \textbf{51} (2019), 147 [arXiv:1907.08284
[astro-ph.IM]].
%42 citations counted in INSPIRE as of 26 Oct 2021




%\cite{Hild:2010id}
\bibitem{Hild:2010id}
S.~Hild, M.~Abernathy, F.~Acernese, P.~Amaro-Seoane, N.~Andersson,
K.~Arun, F.~Barone, B.~Barr, M.~Barsuglia and M.~Beker, \textit{et
al.}
%``Sensitivity Studies for Third-Generation Gravitational Wave Observatories,''
Class. Quant. Grav. \textbf{28} (2011), 094013
doi:10.1088/0264-9381/28/9/094013 [arXiv:1012.0908 [gr-qc]].
%353 citations counted in INSPIRE as of 28 Apr 2021




%\cite{Baker:2019nia}
\bibitem{Baker:2019nia}
J.~Baker, J.~Bellovary, P.~L.~Bender, E.~Berti, R.~Caldwell,
J.~Camp, J.~W.~Conklin, N.~Cornish, C.~Cutler and R.~DeRosa,
\textit{et al.}
%``The Laser Interferometer Space Antenna: Unveiling the Millihertz Gravitational Wave Sky,''
[arXiv:1907.06482 [astro-ph.IM]].
%39 citations counted in INSPIRE as of 28 Apr 2021


%\cite{Smith:2019wny}
\bibitem{Smith:2019wny}
T.~L.~Smith and R.~Caldwell,
%``LISA for Cosmologists: Calculating the Signal-to-Noise Ratio for Stochastic and Deterministic Sources,''
Phys. Rev. D \textbf{100} (2019) no.10, 104055
doi:10.1103/PhysRevD.100.104055 [arXiv:1908.00546 [astro-ph.CO]].
%24 citations counted in INSPIRE as of 28 Apr 2021


%\cite{Crowder:2005nr}
\bibitem{Crowder:2005nr}
J.~Crowder and N.~J.~Cornish,
%``Beyond LISA: Exploring future gravitational wave missions,''
Phys. Rev. D \textbf{72} (2005), 083005
doi:10.1103/PhysRevD.72.083005 [arXiv:gr-qc/0506015 [gr-qc]].
%242 citations counted in INSPIRE as of 28 Apr 2021


%\cite{Smith:2016jqs}
\bibitem{Smith:2016jqs}
T.~L.~Smith and R.~Caldwell,
%``Sensitivity to a Frequency-Dependent Circular Polarization in an Isotropic Stochastic Gravitational Wave Background,''
Phys. Rev. D \textbf{95} (2017) no.4, 044036
doi:10.1103/PhysRevD.95.044036 [arXiv:1609.05901 [gr-qc]].
%36 citations counted in INSPIRE as of 28 Apr 2021



%\cite{Seto:2001qf}
\bibitem{Seto:2001qf}
N.~Seto, S.~Kawamura and T.~Nakamura,
%``Possibility of direct measurement of the acceleration of the universe using 0.1-Hz band laser interferometer gravitational wave antenna in space,''
Phys. Rev. Lett. \textbf{87} (2001), 221103
doi:10.1103/PhysRevLett.87.221103 [arXiv:astro-ph/0108011
[astro-ph]].
%468 citations counted in INSPIRE as of 28 Apr 2021


%\cite{Kawamura:2020pcg}
\bibitem{Kawamura:2020pcg}
S.~Kawamura, M.~Ando, N.~Seto, S.~Sato, M.~Musha, I.~Kawano,
J.~Yokoyama, T.~Tanaka, K.~Ioka and T.~Akutsu, \textit{et al.}
%``Current status of space gravitational wave antenna DECIGO and B-DECIGO,''
[arXiv:2006.13545 [gr-qc]].
%26 citations counted in INSPIRE as of 28 Apr 2021



%\cite{Bull:2018lat}
\bibitem{Bull:2018lat}
A.~Weltman, P.~Bull, S.~Camera, K.~Kelley, H.~Padmanabhan,
J.~Pritchard, A.~Raccanelli, S.~Riemer-S\o{}rensen, L.~Shao and
S.~Andrianomena, \textit{et al.}
%``Fundamental physics with the Square Kilometre Array,''
Publ. Astron. Soc. Austral. \textbf{37} (2020), e002
doi:10.1017/pasa.2019.42 [arXiv:1810.02680 [astro-ph.CO]].
%82 citations counted in INSPIRE as of 28 Apr 2021




%\cite{LISACosmologyWorkingGroup:2022jok}
\bibitem{LISACosmologyWorkingGroup:2022jok}
P.~Auclair \textit{et al.} [LISA Cosmology Working Group],
%``Cosmology with the Laser Interferometer Space Antenna,''
[arXiv:2204.05434 [astro-ph.CO]].
%6 citations counted in INSPIRE as of 17 May 2022











\bibitem{NANOGrav:2023gor}

%\cite{NANOGrav:2023gor}
%\bibitem{NANOGrav:2023gor}
G.~Agazie \textit{et al.} [NANOGrav],
%``The NANOGrav 15-year Data Set: Evidence for a Gravitational-Wave Background,''
doi:10.3847/2041-8213/acdac6 [arXiv:2306.16213 [astro-ph.HE]].
%5 citations counted in INSPIRE as of 29 Jun 2023




%EPTA

%\cite{Antoniadis:2023ott}
\bibitem{Antoniadis:2023ott}
J.~Antoniadis, P.~Arumugam, S.~Arumugam, S.~Babak, M.~Bagchi,
A.~S.~B.~Nielsen, C.~G.~Bassa, A.~Bathula, A.~Berthereau and
M.~Bonetti, \textit{et al.}
%``The second data release from the European Pulsar Timing Array III. Search for gravitational wave signals,''
[arXiv:2306.16214 [astro-ph.HE]].
%0 citations counted in INSPIRE as of 29 Jun 2023



%\cite{Reardon:2023gzh}
\bibitem{Reardon:2023gzh}
D.~J.~Reardon, A.~Zic, R.~M.~Shannon, G.~B.~Hobbs, M.~Bailes,
V.~Di Marco, A.~Kapur, A.~F.~Rogers, E.~Thrane and J.~Askew,
\textit{et al.}
%``Search for an isotropic gravitational-wave background with the Parkes Pulsar Timing Array,''
doi:10.3847/2041-8213/acdd02 [arXiv:2306.16215 [astro-ph.HE]].
%0 citations counted in INSPIRE as of 29 Jun 2023


%\cite{Xu:2023wog}
\bibitem{Xu:2023wog}
H.~Xu, S.~Chen, Y.~Guo, J.~Jiang, B.~Wang, J.~Xu, Z.~Xue,
R.~N.~Caballero, J.~Yuan and Y.~Xu, \textit{et al.}
%``Searching for the nano-Hertz stochastic gravitational wave background with the Chinese Pulsar Timing Array Data Release I,''
doi:10.1088/1674-4527/acdfa5 [arXiv:2306.16216 [astro-ph.HE]].
%0 citations counted in INSPIRE as of 29 Jun 2023



\bibitem{sunnynew}
S.~Vagnozzi,
%``Inflationary interpretation of the stochastic gravitational wave background signal detected by pulsar timing array experiments,''
JHEAp \textbf{39} (2023), 81-98 doi:10.1016/j.jheap.2023.07.001
[arXiv:2306.16912 [astro-ph.CO]].
%59 citations counted in INSPIRE as of 06 Sep 2023




%\cite{Oikonomou:2023qfz}
\bibitem{Oikonomou:2023qfz}
V.~K.~Oikonomou,
%``Flat energy spectrum of primordial gravitational waves versus peaks and the NANOGrav 2023 observation,''
Phys. Rev. D \textbf{108} (2023) no.4, 043516
doi:10.1103/PhysRevD.108.043516 [arXiv:2306.17351 [astro-ph.CO]].
%14 citations counted in INSPIRE as of 29 Aug 2023

%\cite{Cai:2023dls}
\bibitem{Cai:2023dls}
Y.~F.~Cai, X.~C.~He, X.~Ma, S.~F.~Yan and G.~W.~Yuan,
%``Limits on scalar-induced gravitational waves from the stochastic background by pulsar timing array observations,''
[arXiv:2306.17822 [gr-qc]].
%0 citations counted in INSPIRE as of 04 Jul 2023


%\cite{Han:2023olf}
\bibitem{Han:2023olf}
C.~Han, K.~P.~Xie, J.~M.~Yang and M.~Zhang,
%``Self-interacting dark matter implied by nano-Hertz gravitational waves,''
[arXiv:2306.16966 [hep-ph]].
%3 citations counted in INSPIRE as of 04 Jul 2023



%\cite{Guo:2023hyp}
\bibitem{Guo:2023hyp}
S.~Y.~Guo, M.~Khlopov, X.~Liu, L.~Wu, Y.~Wu and B.~Zhu,
%``Footprints of Axion-Like Particle in Pulsar Timing Array Data and JWST Observations,''
[arXiv:2306.17022 [hep-ph]].
%3 citations counted in INSPIRE as of 04 Jul 2023


%\cite{Yang:2023aak}
\bibitem{Yang:2023aak}
J.~Yang, N.~Xie and F.~P.~Huang,
%``Nano-Hertz stochastic gravitational wave background as hints of ultralight axion particles,''
[arXiv:2306.17113 [hep-ph]].
%3 citations counted in INSPIRE as of 04 Jul 2023


%\cite{Addazi:2023jvg}
\bibitem{Addazi:2023jvg}
A.~Addazi, Y.~F.~Cai, A.~Marciano and L.~Visinelli,
%``Have pulsar timing array methods detected a cosmological phase transition?,''
[arXiv:2306.17205 [astro-ph.CO]].
%0 citations counted in INSPIRE as of 04 Jul 2023


%\cite{Li:2023bxy}
\bibitem{Li:2023bxy}
S.~P.~Li and K.~P.~Xie,
%``Collider test of nano-Hertz gravitational waves from pulsar timing arrays,''
Phys. Rev. D \textbf{108} (2023) no.5, 055018
doi:10.1103/PhysRevD.108.055018 [arXiv:2307.01086 [hep-ph]].
%25 citations counted in INSPIRE as of 20 Sep 2023


%\cite{Niu:2023bsr}
\bibitem{Niu:2023bsr}
X.~Niu and M.~H.~Rahat,
%``NANOGrav signal from axion inflation,''
[arXiv:2307.01192 [hep-ph]].
%0 citations counted in INSPIRE as of 04 Jul 2023



%\cite{Yang:2023qlf}
\bibitem{Yang:2023qlf}
A.~Yang, J.~Ma, S.~Jiang and F.~P.~Huang,
%``Implication of nano-Hertz stochastic gravitational wave on dynamical dark matter through a first-order phase transition,''
[arXiv:2306.17827 [hep-ph]].
%10 citations counted in INSPIRE as of 13 Jul 2023




%\cite{Datta:2023vbs}
\bibitem{Datta:2023vbs}
S.~Datta,
%``Inflationary gravitational waves, pulsar timing data and low-scale-leptogenesis,''
[arXiv:2307.00646 [hep-ph]].
%5 citations counted in INSPIRE as of 13 Jul 2023

%\cite{Du:2023qvj}
\bibitem{Du:2023qvj}
X.~K.~Du, M.~X.~Huang, F.~Wang and Y.~K.~Zhang,
%``Did the nHZ Gravitational Waves Signatures Observed By NANOGrav Indicate Multiple Sector SUSY Breaking?,''
[arXiv:2307.02938 [hep-ph]].
%5 citations counted in INSPIRE as of 14 Jul 2023





%\cite{Salvio:2023ynn}
\bibitem{Salvio:2023ynn}
A.~Salvio,
%``Supercooling in Radiative Symmetry Breaking: Theory Extensions, Gravitational Wave Detection and Primordial Black Holes,''
[arXiv:2307.04694 [hep-ph]].
%1 citations counted in INSPIRE as of 14 Jul 2023

%\cite{Yi:2023mbm}
\bibitem{Yi:2023mbm}
Z.~Yi, Q.~Gao, Y.~Gong, Y.~Wang and F.~Zhang,
%``The waveform of the scalar induced gravitational waves in light of Pulsar Timing Array data,''
[arXiv:2307.02467 [gr-qc]].
%7 citations counted in INSPIRE as of 14 Jul 2023


%\cite{You:2023rmn}
\bibitem{You:2023rmn}
Z.~Q.~You, Z.~Yi and Y.~Wu,
%``Constraints on primordial curvature power spectrum with pulsar timing arrays,''
[arXiv:2307.04419 [gr-qc]].
%0 citations counted in INSPIRE as of 14 Jul 2023


%\cite{Wang:2023div}
\bibitem{Wang:2023div}
S.~Wang and Z.~C.~Zhao,
%``Unveiling the Graviton Mass Bounds through Analysis of 2023 Pulsar Timing Array Datasets,''
[arXiv:2307.04680 [astro-ph.HE]].
%0 citations counted in INSPIRE as of 15 Jul 2023



%\cite{Figueroa:2023zhu}
\bibitem{Figueroa:2023zhu}
D.~G.~Figueroa, M.~Pieroni, A.~Ricciardone and P.~Simakachorn,
%``Cosmological Background Interpretation of Pulsar Timing Array Data,''
[arXiv:2307.02399 [astro-ph.CO]].
%14 citations counted in INSPIRE as of 20 Jul 2023


%\cite{Choudhury:2023kam}
\bibitem{Choudhury:2023kam}
S.~Choudhury,
%``Single field inflation in the light of NANOGrav 15-year Data: Quintessential interpretation of blue tilted tensor spectrum through Non-Bunch Davies initial condition,''
[arXiv:2307.03249 [astro-ph.CO]].
%3 citations counted in INSPIRE as of 20 Jul 2023


%\cite{HosseiniMansoori:2023mqh}
\bibitem{HosseiniMansoori:2023mqh}
S.~A.~Hosseini Mansoori, F.~Felegray, A.~Talebian and M.~Sami,
%``PBHs and GWs from $\mathbb{T}^2$-inflation and NANOGrav 15-year data,''
[arXiv:2307.06757 [astro-ph.CO]].
%1 citations counted in INSPIRE as of 20 Jul 2023


%\cite{Ge:2023rce}
\bibitem{Ge:2023rce}
S.~Ge,
%``Stochastic gravitational wave background: birth from axionic string-wall death,''
[arXiv:2307.08185 [gr-qc]].
%1 citations counted in INSPIRE as of 30 Jul 2023

%\cite{Bian:2023dnv}
\bibitem{Bian:2023dnv}
L.~Bian, S.~Ge, J.~Shu, B.~Wang, X.~Y.~Yang and J.~Zong,
%``Gravitational wave sources for Pulsar Timing Arrays,''
[arXiv:2307.02376 [astro-ph.HE]].
%10 citations counted in INSPIRE as of 30 Jul 2023

%\cite{Kawasaki:2023rfx}
\bibitem{Kawasaki:2023rfx}
M.~Kawasaki and K.~Murai,
%``Enhancement of gravitational waves at Q-ball decay including non-linear density perturbations,''
[arXiv:2308.13134 [astro-ph.CO]].
%1 citations counted in INSPIRE as of 29 Aug 2023





%\cite{Yi:2023tdk}
\bibitem{Yi:2023tdk}
Z.~Yi, Z.~Q.~You and Y.~Wu,
%``Model-independent reconstruction of the primordial curvature power spectrum from PTA data,''
[arXiv:2308.05632 [astro-ph.CO]].
%3 citations counted in INSPIRE as of 29 Aug 2023


%\cite{An:2023jxf}
\bibitem{An:2023jxf}
H.~An, B.~Su, H.~Tai, L.~T.~Wang and C.~Yang,
%``Phase transition during inflation and the gravitational wave signal at pulsar timing arrays,''
[arXiv:2308.00070 [astro-ph.CO]].
%6 citations counted in INSPIRE as of 29 Aug 2023


%\cite{Zhang:2023nrs}
\bibitem{Zhang:2023nrs}
Z.~Zhang, C.~Cai, Y.~H.~Su, S.~Wang, Z.~H.~Yu and H.~H.~Zhang,
%``Nano-Hertz gravitational waves from collapsing domain walls associated with freeze-in dark matter in light of pulsar timing array observations,''
[arXiv:2307.11495 [hep-ph]].
%14 citations counted in INSPIRE as of 29 Aug 2023

%\cite{DiBari:2023upq}
\bibitem{DiBari:2023upq}
P.~Di Bari and M.~H.~Rahat,
%``The split majoron model confronts the NANOGrav signal,''
[arXiv:2307.03184 [hep-ph]].
%14 citations counted in INSPIRE as of 29 Aug 2023


%\cite{Jiang:2023qbm}
\bibitem{Jiang:2023qbm}
S.~Jiang, A.~Yang, J.~Ma and F.~P.~Huang,
%``Implication of nano-Hertz stochastic gravitational wave on dynamical dark matter through a first-order phase transition,''
[arXiv:2306.17827 [hep-ph]].
%17 citations counted in INSPIRE as of 29 Aug 2023


%\cite{Bhattacharya:2023ysp}
\bibitem{Bhattacharya:2023ysp}
G.~Bhattacharya, S.~Choudhury, K.~Dey, S.~Ghosh, A.~Karde and
N.~S.~Mishra,
%``Evading no-go for PBH formation and production of SIGWs using Multiple Sharp Transitions in EFT of single field inflation,''
[arXiv:2309.00973 [astro-ph.CO]].
%0 citations counted in INSPIRE as of 06 Sep 2023




%\cite{Choudhury:2023hfm}
\bibitem{Choudhury:2023hfm}
S.~Choudhury, A.~Karde, S.~Panda and M.~Sami,
%``Scalar induced gravity waves from ultra slow-roll Galileon inflation,''
[arXiv:2308.09273 [astro-ph.CO]].
%3 citations counted in INSPIRE as of 06 Sep 2023



%\cite{Bringmann:2023opz}
\bibitem{Bringmann:2023opz}
T.~Bringmann, P.~F.~Depta, T.~Konstandin, K.~Schmidt-Hoberg and
C.~Tasillo,
%``Does NANOGrav observe a dark sector phase transition?,''
[arXiv:2306.09411 [astro-ph.CO]].
%0 citations counted in INSPIRE as of 22 Jun 2023

%\cite{Choudhury:2023hvf}
\bibitem{Choudhury:2023hvf}
S.~Choudhury, S.~Panda and M.~Sami,
%``Galileon inflation evades the no-go for PBH formation in the single-field framework,''
JCAP \textbf{08} (2023), 078 doi:10.1088/1475-7516/2023/08/078
[arXiv:2304.04065 [astro-ph.CO]].
%23 citations counted in INSPIRE as of 12 Sep 2023


%\cite{Choudhury:2023kdb}
\bibitem{Choudhury:2023kdb}
S.~Choudhury, A.~Karde, S.~Panda and M.~Sami,
%``Primordial non-Gaussianity from ultra slow-roll Galileon inflation,''
[arXiv:2306.12334 [astro-ph.CO]].
%9 citations counted in INSPIRE as of 12 Sep 2023


%\cite{Huang:2023chx}
\bibitem{Huang:2023chx}
H.~L.~Huang, Y.~Cai, J.~Q.~Jiang, J.~Zhang and Y.~S.~Piao,
%``Supermassive primordial black holes in multiverse: for nano-Hertz gravitational wave and high-redshift JWST galaxies,''
[arXiv:2306.17577 [gr-qc]].
%20 citations counted in INSPIRE as of 12 Sep 2023


%\cite{Jiang:2023gfe}
\bibitem{Jiang:2023gfe}
J.~Q.~Jiang, Y.~Cai, G.~Ye and Y.~S.~Piao,
%``Broken blue-tilted inflationary gravitational waves: a joint analysis of NANOGrav 15-year and BICEP/Keck 2018 data,''
[arXiv:2307.15547 [astro-ph.CO]].
%8 citations counted in INSPIRE as of 12 Sep 2023

%\cite{Zhu:2023lbf}
\bibitem{Zhu:2023lbf}
M.~Zhu, G.~Ye and Y.~Cai,
%``Pulsar timing array observations as a possible hint for nonsingular cosmology,''
[arXiv:2307.16211 [astro-ph.CO]].
%3 citations counted in INSPIRE as of 12 Sep 2023

%\cite{Ben-Dayan:2023lwd}
\bibitem{Ben-Dayan:2023lwd}
I.~Ben-Dayan, U.~Kumar, U.~Thattarampilly and A.~Verma,
%``Probing The Early Universe Cosmology With NANOGrav: Possibilities and Limitations,''
[arXiv:2307.15123 [astro-ph.CO]].
%7 citations counted in INSPIRE as of 12 Sep 2023




%\cite{Franciolini:2023pbf}
\bibitem{Franciolini:2023pbf}
G.~Franciolini, A.~Iovino, Junior., V.~Vaskonen and H.~Veermae,
%``The recent gravitational wave observation by pulsar timing arrays and primordial black holes: the importance of non-gaussianities,''
[arXiv:2306.17149 [astro-ph.CO]].
%64 citations counted in INSPIRE as of 12 Sep 2023


%\cite{Ellis:2023oxs}
\bibitem{Ellis:2023oxs}
J.~Ellis, M.~Fairbairn, G.~Franciolini, G.~H\"utsi, A.~Iovino,
M.~Lewicki, M.~Raidal, J.~Urrutia, V.~Vaskonen and H.~Veerm\"ae,
%``What is the source of the PTA GW signal?,''
[arXiv:2308.08546 [astro-ph.CO]].
%9 citations counted in INSPIRE as of 12 Sep 2023



%\cite{Liu:2023ymk}
\bibitem{Liu:2023ymk}
L.~Liu, Z.~C.~Chen and Q.~G.~Huang,
%``Implications for the non-Gaussianity of curvature perturbation from pulsar timing arrays,''
[arXiv:2307.01102 [astro-ph.CO]].
%39 citations counted in INSPIRE as of 14 Sep 2023

%\cite{Liu:2023pau}
\bibitem{Liu:2023pau}
L.~Liu, Z.~C.~Chen and Q.~G.~Huang,
%``Probing the equation of state of the early Universe with pulsar timing arrays,''
[arXiv:2307.14911 [astro-ph.CO]].
%14 citations counted in INSPIRE as of 14 Sep 2023


%\cite{Madge:2023cak}
\bibitem{Madge:2023cak}
E.~Madge, E.~Morgante, C.~Puchades-Ib\'a\~nez, N.~Ramberg,
W.~Ratzinger, S.~Schenk and P.~Schwaller,
%``Primordial gravitational waves in the nano-Hertz regime and PTA data -- towards solving the GW inverse problem,''
[arXiv:2306.14856 [hep-ph]].
%40 citations counted in INSPIRE as of 14 Sep 2023


%\cite{Huang:2023zvs}
\bibitem{Huang:2023zvs}
M.~X.~Huang, F.~Wang and Y.~K.~Zhang,
%``The Interplay Between the Muon $g-2$ Anomaly and the PTA nHZ Gravitational Waves from Domain Walls in NMSSM,''
[arXiv:2309.06378 [hep-ph]].
%0 citations counted in INSPIRE as of 14 Sep 2023


%\cite{Fu:2023aab}
\bibitem{Fu:2023aab}
C.~Fu, J.~Liu, X.~Y.~Yang, W.~W.~Yu and Y.~Zhang,
%``Explaining Pulsar Timing Array Observations with Primordial Gravitational Waves in Parity-Violating Gravity,''
[arXiv:2308.15329 [astro-ph.CO]].
%2 citations counted in INSPIRE as of 14 Sep 2023


%\cite{Maji:2023fhv}
\bibitem{Maji:2023fhv}
R.~Maji and W.~I.~Park,
%``Supersymmetric $U(1)_{B-L}$ flat direction and NANOGrav 15 year data,''
[arXiv:2308.11439 [hep-ph]].
%3 citations counted in INSPIRE as of 14 Sep 2023

%\cite{Gangopadhyay:2023qjr}
\bibitem{Gangopadhyay:2023qjr}
M.~R.~Gangopadhyay, V.~V.~Godithi, K.~Ichiki, R.~Inui, T.~Kajino,
A.~Manusankar, G.~J.~Mathews and Yogesh,
%``Is NanoGRAV signals pointing towards resonant particle creation during inflation?,''
[arXiv:2309.03101 [astro-ph.CO]].
%1 citations counted in INSPIRE as of 16 Sep 2023


%\cite{Wang:2023sij}
\bibitem{Wang:2023sij}
S.~Wang, Z.~C.~Zhao and Q.~H.~Zhu,
%``Constraints On Scalar-Induced Gravitational Waves Up To Third Order From Joint Analysis of BBN, CMB, And PTA Data,''
[arXiv:2307.03095 [astro-ph.CO]].
%41 citations counted in INSPIRE as of 18 Feb 2024

%\cite{Wang:2023ost}
\bibitem{Wang:2023ost}
S.~Wang, Z.~C.~Zhao, J.~P.~Li and Q.~H.~Zhu,
%``Implications of Pulsar Timing Array Data for Scalar-Induced Gravitational Waves and Primordial Black Holes: Primordial Non-Gaussianity $f_{\mathrm{NL}}$ Considered,''
[arXiv:2307.00572 [astro-ph.CO]].
%65 citations counted in INSPIRE as of 18 Feb 2024


%\cite{Schwaller:2015tja}
\bibitem{Schwaller:2015tja}
P.~Schwaller,
%``Gravitational Waves from a Dark Phase Transition,''
Phys. Rev. Lett. \textbf{115} (2015) no.18, 181101
doi:10.1103/PhysRevLett.115.181101 [arXiv:1504.07263 [hep-ph]].
%240 citations counted in INSPIRE as of 04 Jul 2023




%\cite{Ratzinger:2020koh}
\bibitem{Ratzinger:2020koh}
W.~Ratzinger and P.~Schwaller,
%``Whispers from the dark side: Confronting light new physics with NANOGrav data,''
SciPost Phys. \textbf{10} (2021) no.2, 047
doi:10.21468/SciPostPhys.10.2.047 [arXiv:2009.11875
[astro-ph.CO]].
%86 citations counted in INSPIRE as of 04 Jul 2023



%\cite{Ashoorioon:2022raz}
\bibitem{Ashoorioon:2022raz}
A.~Ashoorioon, K.~Rezazadeh and A.~Rostami,
%``NANOGrav signal from the end of inflation and the LIGO mass and heavier primordial black holes,''
Phys. Lett. B \textbf{835} (2022), 137542
doi:10.1016/j.physletb.2022.137542 [arXiv:2202.01131
[astro-ph.CO]].
%39 citations counted in INSPIRE as of 04 Jul 2023

%\cite{Choudhury:2023vuj}
\bibitem{Choudhury:2023vuj}
S.~Choudhury, M.~R.~Gangopadhyay and M.~Sami,
%``No-go for the formation of heavy mass Primordial Black Holes in Single Field Inflation,''
[arXiv:2301.10000 [astro-ph.CO]].
%36 citations counted in INSPIRE as of 20 Jul 2023

%\cite{Choudhury:2023jlt}
\bibitem{Choudhury:2023jlt}
S.~Choudhury, S.~Panda and M.~Sami,
%``No-go for PBH formation in EFT of single field inflation,''
[arXiv:2302.05655 [astro-ph.CO]].
%32 citations counted in INSPIRE as of 20 Jul 2023


%\cite{Choudhury:2023rks}
\bibitem{Choudhury:2023rks}
S.~Choudhury, S.~Panda and M.~Sami,
%``Quantum loop effects on the power spectrum and constraints on primordial black holes,''
[arXiv:2303.06066 [astro-ph.CO]].
%29 citations counted in INSPIRE as of 20 Jul 2023


%\cite{Bian:2022qbh}
\bibitem{Bian:2022qbh}
L.~Bian, S.~Ge, C.~Li, J.~Shu and J.~Zong,
%``Domain Wall Network: A Dual Solution for Gravitational Waves and Hubble Tension?,''
[arXiv:2212.07871 [hep-ph]].
%9 citations counted in INSPIRE as of 30 Jul 2023






%\cite{Machado:2018nqk}
\bibitem{Machado:2018nqk}
C.~S.~Machado, W.~Ratzinger, P.~Schwaller and B.~A.~Stefanek,
%``Audible Axions,''
JHEP \textbf{01} (2019), 053 doi:10.1007/JHEP01(2019)053
[arXiv:1811.01950 [hep-ph]].
%54 citations counted in INSPIRE as of 04 Jul 2023





%\cite{Regimbau:2022mdu}
\bibitem{Regimbau:2022mdu}
T.~Regimbau,
%``The Quest for the Astrophysical Gravitational-Wave Background with Terrestrial Detectors,''
Symmetry \textbf{14} (2022) no.2, 270 doi:10.3390/sym14020270
%11 citations counted in INSPIRE as of 05 Jul 2023








%\cite{Benetti:2021uea}
\bibitem{Benetti:2021uea}
M.~Benetti, L.~L.~Graef and S.~Vagnozzi,
%``Primordial gravitational waves from NANOGrav: A broken power-law approach,''
Phys. Rev. D \textbf{105} (2022) no.4, 043520
doi:10.1103/PhysRevD.105.043520 [arXiv:2111.04758 [astro-ph.CO]].
%38 citations counted in INSPIRE as of 06 Jun 2023




%\cite{Vagnozzi:2020gtf}
\bibitem{Vagnozzi:2020gtf}
S.~Vagnozzi,
%``Implications of the NANOGrav results for inflation,''
Mon. Not. Roy. Astron. Soc. \textbf{502} (2021) no.1, L11-L15
doi:10.1093/mnrasl/slaa203 [arXiv:2009.13432 [astro-ph.CO]].
%92 citations counted in INSPIRE as of 06 Jun 2023






%\cite{Kamali:2020drm}
\bibitem{Kamali:2020drm}
V.~Kamali and R.~Brandenberger,
%``Creating spatial flatness by combining string gas cosmology and power law inflation,''
Phys. Rev. D \textbf{101} (2020) no.10, 103512
doi:10.1103/PhysRevD.101.103512 [arXiv:2002.09771 [hep-th]].
%3 citations counted in INSPIRE as of 20 Jul 2023



%\cite{Brandenberger:2015kga}
\bibitem{Brandenberger:2015kga}
R.~H.~Brandenberger,
%``String Gas Cosmology after Planck,''
Class. Quant. Grav. \textbf{32} (2015) no.23, 234002
doi:10.1088/0264-9381/32/23/234002 [arXiv:1505.02381 [hep-th]].
%34 citations counted in INSPIRE as of 20 Jul 2023



%\cite{Brandenberger:2006pr}
\bibitem{Brandenberger:2006pr}
R.~H.~Brandenberger, S.~Kanno, J.~Soda, D.~A.~Easson, J.~Khoury,
P.~Martineau, A.~Nayeri and S.~P.~Patil,
%``More on the spectrum of perturbations in string gas cosmology,''
JCAP \textbf{11} (2006), 009 doi:10.1088/1475-7516/2006/11/009
[arXiv:hep-th/0608186 [hep-th]].
%70 citations counted in INSPIRE as of 20 Jul 2023


%\cite{Ashtekar:2011ni}
\bibitem{Ashtekar:2011ni}
A.~Ashtekar and P.~Singh,
%``Loop Quantum Cosmology: A Status Report,''
Class. Quant. Grav. \textbf{28} (2011), 213001
doi:10.1088/0264-9381/28/21/213001 [arXiv:1108.0893 [gr-qc]].
%927 citations counted in INSPIRE as of 06 Sep 2023





%\cite{Bojowald:2011iq}
\bibitem{Bojowald:2011iq}
M.~Bojowald, G.~Calcagni and S.~Tsujikawa,
%``Observational test of inflation in loop quantum cosmology,''
JCAP \textbf{11} (2011), 046 doi:10.1088/1475-7516/2011/11/046
[arXiv:1107.1540 [gr-qc]].
%86 citations counted in INSPIRE as of 06 Sep 2023


%\cite{Mielczarek:2009vi}
\bibitem{Mielczarek:2009vi}
J.~Mielczarek,
%``Tensor power spectrum with holonomy corrections in LQC,''
Phys. Rev. D \textbf{79} (2009), 123520
doi:10.1103/PhysRevD.79.123520 [arXiv:0902.2490 [gr-qc]].
%34 citations counted in INSPIRE as of 06 Sep 2023


%\cite{Bojowald:2008ik}
\bibitem{Bojowald:2008ik}
M.~Bojowald,
%``Consistent Loop Quantum Cosmology,''
Class. Quant. Grav. \textbf{26} (2009), 075020
doi:10.1088/0264-9381/26/7/075020 [arXiv:0811.4129 [gr-qc]].
%103 citations counted in INSPIRE as of 06 Sep 2023







%\cite{Calcagni:2020tvw}
\bibitem{Calcagni:2020tvw}
G.~Calcagni and S.~Kuroyanagi,
%``Stochastic gravitational-wave background in quantum gravity,''
JCAP \textbf{03} (2021), 019 doi:10.1088/1475-7516/2021/03/019
[arXiv:2012.00170 [gr-qc]].
%29 citations counted in INSPIRE as of 20 Jul 2023



%\cite{Koshelev:2020foq}
\bibitem{Koshelev:2020foq}
A.~S.~Koshelev, K.~Sravan Kumar, A.~Mazumdar and
A.~A.~Starobinsky,
%``Non-Gaussianities and tensor-to-scalar ratio in non-local R$^{2}$-like inflation,''
JHEP \textbf{06} (2020), 152 doi:10.1007/JHEP06(2020)152
[arXiv:2003.00629 [hep-th]].
%38 citations counted in INSPIRE as of 20 Jul 2023


%\cite{Koshelev:2017tvv}
\bibitem{Koshelev:2017tvv}
A.~S.~Koshelev, K.~Sravan Kumar and A.~A.~Starobinsky,
%``$R^2$ inflation to probe non-perturbative quantum gravity,''
JHEP \textbf{03} (2018), 071 doi:10.1007/JHEP03(2018)071
[arXiv:1711.08864 [hep-th]].
%73 citations counted in INSPIRE as of 20 Jul 2023


%\cite{Baumgart:2021ptt}
\bibitem{Baumgart:2021ptt}
M.~Baumgart, J.~J.~Heckman and L.~Thomas,
%``CFTs blueshift tensor fluctuations universally,''
JCAP \textbf{07} (2022) no.07, 034
doi:10.1088/1475-7516/2022/07/034 [arXiv:2109.08166 [hep-ph]].
%6 citations counted in INSPIRE as of 15 Aug 2023










%\cite{Co:2021lkc}
\bibitem{Co:2021lkc}
R.~T.~Co, D.~Dunsky, N.~Fernandez, A.~Ghalsasi, L.~J.~Hall,
K.~Harigaya and J.~Shelton,
%``Gravitational wave and CMB probes of axion kination,''
JHEP \textbf{09} (2022), 116 doi:10.1007/JHEP09(2022)116
[arXiv:2108.09299 [hep-ph]].
%42 citations counted in INSPIRE as of 06 Sep 2023


%\cite{Gouttenoire:2021jhk}
\bibitem{Gouttenoire:2021jhk}
Y.~Gouttenoire, G.~Servant and P.~Simakachorn,
%``Kination cosmology from scalar fields and gravitational-wave signatures,''
[arXiv:2111.01150 [hep-ph]].
%39 citations counted in INSPIRE as of 06 Jun 2023


%\cite{Giovannini:1998bp}
\bibitem{Giovannini:1998bp}
M.~Giovannini,
%``Gravitational waves constraints on postinflationary phases stiffer than radiation,''
Phys. Rev. D \textbf{58} (1998), 083504
doi:10.1103/PhysRevD.58.083504 [arXiv:hep-ph/9806329 [hep-ph]].
%138 citations counted in INSPIRE as of 25 Mar 2023


%\cite{Oikonomou:2023bah}
\bibitem{Oikonomou:2023bah}
V.~K.~Oikonomou,
%``Effects of the axion through the Higgs portal on primordial gravitational waves during the electroweak breaking,''
Phys. Rev. D \textbf{107} (2023) no.6, 064071
doi:10.1103/PhysRevD.107.064071 [arXiv:2303.05889 [hep-ph]].
%8 citations counted in INSPIRE as of 06 Sep 2023




%\cite{Ford:1986sy}
\bibitem{Ford:1986sy}
L.~H.~Ford,
%``Gravitational Particle Creation and Inflation,''
Phys. Rev. D \textbf{35} (1987), 2955 doi:10.1103/PhysRevD.35.2955
%518 citations counted in INSPIRE as of 18 Mar 2023


%\cite{Kamionkowski:1990ni}
\bibitem{Kamionkowski:1990ni}
M.~Kamionkowski and M.~S.~Turner,
%``THERMAL RELICS: DO WE KNOW THEIR ABUNDANCES?,''
Phys. Rev. D \textbf{42} (1990), 3310-3320
doi:10.1103/PhysRevD.42.3310
%223 citations counted in INSPIRE as of 18 Mar 2023




%\cite{Grin:2007yg}
\bibitem{Grin:2007yg}
D.~Grin, T.~L.~Smith and M.~Kamionkowski,
%``Axion constraints in non-standard thermal histories,''
Phys. Rev. D \textbf{77} (2008), 085020
doi:10.1103/PhysRevD.77.085020 [arXiv:0711.1352 [astro-ph]].
%62 citations counted in INSPIRE as of 18 Mar 2023



%\cite{Visinelli:2009kt}
\bibitem{Visinelli:2009kt}
L.~Visinelli and P.~Gondolo,
%``Axion cold dark matter in non-standard cosmologies,''
Phys. Rev. D \textbf{81} (2010), 063508
doi:10.1103/PhysRevD.81.063508 [arXiv:0912.0015 [astro-ph.CO]].
%103 citations counted in INSPIRE as of 18 Mar 2023


%\cite{Giovannini:1999qj}
\bibitem{Giovannini:1999qj}
M.~Giovannini,
%``Spikes in the relic graviton background from quintessential inflation,''
Class. Quant. Grav. \textbf{16} (1999), 2905-2913
doi:10.1088/0264-9381/16/9/308 [arXiv:hep-ph/9903263 [hep-ph]].
%104 citations counted in INSPIRE as of 25 Mar 2023

%\cite{Giovannini:1999bh}
\bibitem{Giovannini:1999bh}
M.~Giovannini,
%``Production and detection of relic gravitons in quintessential inflationary models,''
Phys. Rev. D \textbf{60} (1999), 123511
doi:10.1103/PhysRevD.60.123511 [arXiv:astro-ph/9903004
[astro-ph]].
%205 citations counted in INSPIRE as of 25 Mar 2023





%\cite{Harigaya:2023pmw}
\bibitem{Harigaya:2023pmw}
K.~Harigaya, K.~Inomata and T.~Terada,
%``Induced Gravitational Waves with Kination Era for Recent Pulsar Timing Array Signals,''
[arXiv:2309.00228 [astro-ph.CO]].
%0 citations counted in INSPIRE as of 06 Sep 2023






\bibitem{reviews1}
 S.~Nojiri, S.~D.~Odintsov and V.~K.~Oikonomou,
  %``Modified Gravity Theories on a Nutshell: Inflation, Bounce and Late-time Evolution,''
  Phys.\ Rept.\  {\bf 692} (2017) 1
  %doi:10.1016/j.physrep.2017.06.001
  [arXiv:1705.11098 [gr-qc]].
  %%CITATION = %doi:10.1016/j.physrep.2017.06.001;%%
  %21 citations counted in INSPIRE as of 20 Aug 2017



\bibitem{reviews2}


 S. Capozziello, M. De Laurentis,
   %``Extended Theories of Gravity,''
   Phys.\ Rept.\  {\bf 509}, 167 (2011);\\
   %[arXiv:1108.6266 [gr-qc]].
   %%CITATION = ARXIV:1108.6266;%%
 V.~Faraoni and S.~Capozziello,
  %``Beyond Einstein Gravity : A Survey of Gravitational Theories for Cosmology and Astrophysics,''
  Fundam.\ Theor.\ Phys.\  {\bf 170} (2010).
  %%doi:10.1007/978-94-007-0165-6
  %%CITATION = %doi:10.1007/978-94-007-0165-6;%%
  %64 citations counted in INSPIRE as of 16 Sep 2017



\bibitem{reviews3}
S. Nojiri, S.D. Odintsov,
  %``Introduction to modified gravity and gravitational alternative for dark
  %energy,''
  eConf {\bf C0602061}, 06 (2006)
  [Int.\ J.\ Geom.\ Meth.\ Mod.\ Phys.\  {\bf 4}, 115 (2007)].
  %[arXiv:hep-th/0601213];
  %%CITATION = 00436,4,115;%%


   \bibitem{reviews4}

S. Nojiri, S.D. Odintsov,
   %``Unified cosmic history in modified gravity: from F(R) theory to
   %Lorentz non-invariant models,''
   Phys.\ Rept.\  {\bf 505}, 59 (2011);
   %[arXiv:1011.0544 [gr-qc]].
   %%CITATION = ARXIV:1011.0544;%%




\bibitem{reviews5}

G.~J.~Olmo,
  %``Palatini Approach to Modified Gravity: f(R) Theories and Beyond,''
  Int.\ J.\ Mod.\ Phys.\ D {\bf 20} (2011) 413
  %doi:10.1142/S0218271811018925
  [arXiv:1101.3864 [gr-qc]].
  %%CITATION = %doi:10.1142/S0218271811018925;%%
  %204 citations counted in INSPIRE as of 04 Feb 2018

%\cite{Sebastiani:2016ras}
\bibitem{Sebastiani:2016ras}
L.~Sebastiani, S.~Vagnozzi and R.~Myrzakulov,
%``Mimetic gravity: a review of recent developments and applications to cosmology and astrophysics,''
Adv. High Energy Phys. \textbf{2017} (2017), 3156915
doi:10.1155/2017/3156915 [arXiv:1612.08661 [gr-qc]].
%198 citations counted in INSPIRE as of 08 Jun 2023

%\cite{Odintsov:2023weg}
\bibitem{reviews6}
S.~D.~Odintsov, V.~K.~Oikonomou, I.~Giannakoudi, F.~P.~Fronimos
and E.~C.~Lymperiadou,
%``Recent Advances on Inflation,''
[arXiv:2307.16308 [gr-qc]].
%2 citations counted in INSPIRE as of 29 Aug 2023




%%%%%%%%%%%%%%%%%%%%%%%%%%%%%%%%%%%%%%%%%%%%%%%%%%%%%%%%%%%%%%%%%%%%%%%%%%%%%%%%%%%%%%%%%%%%%%%%%%%%%%%%%%%%%%%%%%%%%%%%%%olga

%\cite{Copeland:1997et}
\bibitem{Copeland:1997et}
E.~J.~Copeland, A.~R.~Liddle and D.~Wands,
%``Exponential potentials and cosmological scaling solutions,''
Phys. Rev. D \textbf{57} (1998), 4686-4690
doi:10.1103/PhysRevD.57.4686 [arXiv:gr-qc/9711068 [gr-qc]].
%1264 citations counted in INSPIRE as of 28 Oct 2023



%\cite{Boehmer:2008av}
\bibitem{Boehmer:2008av}
C.~G.~Boehmer, G.~Caldera-Cabral, R.~Lazkoz and R.~Maartens,
%``Dynamics of dark energy with a coupling to dark matter,''
Phys. Rev. D \textbf{78} (2008), 023505
doi:10.1103/PhysRevD.78.023505 [arXiv:0801.1565 [gr-qc]].
%313 citations counted in INSPIRE as of 28 Oct 2023


%\cite{Yang:2022csz}
\bibitem{Yang:2022csz}
W.~Yang, S.~Pan, O.~Mena and E.~Di Valentino,
%``On the dynamics of a dark sector coupling,''
JHEAp \textbf{40} (2023), 19-40 doi:10.1016/j.jheap.2023.09.001
[arXiv:2209.14816 [astro-ph.CO]].
%11 citations counted in INSPIRE as of 28 Oct 2023


%\cite{Nunes:2022bhn}
\bibitem{Nunes:2022bhn}
R.~C.~Nunes, S.~Vagnozzi, S.~Kumar, E.~Di Valentino and O.~Mena,
%``New tests of dark sector interactions from the full-shape galaxy power spectrum,''
Phys. Rev. D \textbf{105} (2022) no.12, 123506
doi:10.1103/PhysRevD.105.123506 [arXiv:2203.08093 [astro-ph.CO]].
%52 citations counted in INSPIRE as of 28 Oct 2023




%\cite{Gariazzo:2021qtg}
\bibitem{Gariazzo:2021qtg}
S.~Gariazzo, E.~Di Valentino, O.~Mena and R.~C.~Nunes,
%``Late-time interacting cosmologies and the Hubble constant tension,''
Phys. Rev. D \textbf{106} (2022) no.2, 023530
doi:10.1103/PhysRevD.106.023530 [arXiv:2111.03152 [astro-ph.CO]].
%27 citations counted in INSPIRE as of 28 Oct 2023


%\cite{Nunes:2021zzi}
\bibitem{Nunes:2021zzi}
R.~C.~Nunes and E.~Di Valentino,
%``Dark sector interaction and the supernova absolute magnitude tension,''
Phys. Rev. D \textbf{104} (2021) no.6, 063529
doi:10.1103/PhysRevD.104.063529 [arXiv:2107.09151 [astro-ph.CO]].
%42 citations counted in INSPIRE as of 28 Oct 2023


%\cite{Yang:2021flj}
\bibitem{Yang:2021flj}
W.~Yang, E.~Di Valentino, S.~Pan, Y.~Wu and J.~Lu,
%``Dynamical dark energy after Planck CMB final release and $H_0$ tension,''
Mon. Not. Roy. Astron. Soc. \textbf{501} (2021) no.4, 5845-5858
doi:10.1093/mnras/staa3914 [arXiv:2101.02168 [astro-ph.CO]].
%52 citations counted in INSPIRE as of 28 Oct 2023



%\cite{Yang:2019uzo}
\bibitem{Yang:2019uzo}
W.~Yang, O.~Mena, S.~Pan and E.~Di Valentino,
%``Dark sectors with dynamical coupling,''
Phys. Rev. D \textbf{100} (2019) no.8, 083509
doi:10.1103/PhysRevD.100.083509 [arXiv:1906.11697 [astro-ph.CO]].
%67 citations counted in INSPIRE as of 28 Oct 2023



%\cite{Yang:2019qza}
\bibitem{Yang:2019qza}
W.~Yang, S.~Pan, E.~Di Valentino, A.~Paliathanasis and J.~Lu,
%``Challenging bulk viscous unified scenarios with cosmological observations,''
Phys. Rev. D \textbf{100} (2019) no.10, 103518
doi:10.1103/PhysRevD.100.103518 [arXiv:1906.04162 [astro-ph.CO]].
%46 citations counted in INSPIRE as of 28 Oct 2023


%\cite{Yang:2018uae}
\bibitem{Yang:2018uae}
W.~Yang, A.~Mukherjee, E.~Di Valentino and S.~Pan,
%``Interacting dark energy with time varying equation of state and the $H_0$ tension,''
Phys. Rev. D \textbf{98} (2018) no.12, 123527
doi:10.1103/PhysRevD.98.123527 [arXiv:1809.06883 [astro-ph.CO]].
%123 citations counted in INSPIRE as of 28 Oct 2023



%\cite{Appleby:2009uf}
\bibitem{Appleby:2009uf}
S.~A.~Appleby, R.~A.~Battye and A.~A.~Starobinsky,
%``Curing singularities in cosmological evolution of F(R) gravity,''
JCAP \textbf{06} (2010), 005 doi:10.1088/1475-7516/2010/06/005
[arXiv:0909.1737 [astro-ph.CO]].
%233 citations counted in INSPIRE as of 28 Oct 2023


%\cite{Wetterich:1994bg}
\bibitem{Wetterich:1994bg}
C.~Wetterich,
%``The Cosmon model for an asymptotically vanishing time dependent cosmological 'constant',''
Astron. Astrophys. \textbf{301} (1995), 321-328
[arXiv:hep-th/9408025 [hep-th]].
%938 citations counted in INSPIRE as of 28 Oct 2023


%\cite{Planck:2018vyg}
\bibitem{Planck:2018vyg}
N.~Aghanim \textit{et al.} [Planck],
%``Planck 2018 results. VI. Cosmological parameters,''
Astron. Astrophys. \textbf{641} (2020), A6 [erratum: Astron.
Astrophys. \textbf{652} (2021), C4]
doi:10.1051/0004-6361/201833910 [arXiv:1807.06209 [astro-ph.CO]].
%11909 citations counted in INSPIRE as of 28 Oct 2023



%\cite{Oikonomou:2020oex}
\bibitem{Oikonomou:2020oex}
V.~K.~Oikonomou,
%``Rescaled Einstein-Hilbert Gravity from $f(R)$ Gravity: Inflation, Dark Energy and the Swampland Criteria,''
Phys. Rev. D \textbf{103} (2021) no.12, 124028
doi:10.1103/PhysRevD.103.124028 [arXiv:2012.01312 [gr-qc]].
%64 citations counted in INSPIRE as of 28 Oct 2023




%\cite{Hwang:2005hb}
\bibitem{Hwang:2005hb}
J.~c.~Hwang and H.~Noh,
%``Classical evolution and quantum generation in generalized gravity theories including string corrections and tachyon: Unified analyses,''
Phys. Rev. D \textbf{71} (2005), 063536
doi:10.1103/PhysRevD.71.063536 [arXiv:gr-qc/0412126 [gr-qc]].
%212 citations counted in INSPIRE as of 28 Oct 2023


%\cite{Odintsov:2020thl}
\bibitem{Odintsov:2020thl}
S.~D.~Odintsov and V.~K.~Oikonomou,
%``Inflationary attractors in F(R) gravity,''
Phys. Lett. B \textbf{807} (2020), 135576
doi:10.1016/j.physletb.2020.135576 [arXiv:2005.12804 [gr-qc]].
%30 citations counted in INSPIRE as of 28 Oct 2023



%%%%%%%%%%%%%%%%%%%%%%%%%%%%%%%%%%%%%%%%%%%%%%%%%%%%%%%%%%%%%%%%%%%%%%%%%%%%%%%%%%%%%%%%%%%%%%stop check




%\cite{Kamionkowski:2015yta}
\bibitem{Kamionkowski:2015yta}
M.~Kamionkowski and E.~D.~Kovetz,
%``The Quest for B Modes from Inflationary Gravitational Waves,''
Ann. Rev. Astron. Astrophys. \textbf{54} (2016), 227-269
doi:10.1146/annurev-astro-081915-023433 [arXiv:1510.06042
[astro-ph.CO]].
%223 citations counted in INSPIRE as of 10 May 2022




%\cite{Turner:1993vb}
\bibitem{Turner:1993vb}
M.~S.~Turner, M.~J.~White and J.~E.~Lidsey,
%``Tensor perturbations in inflationary models as a probe of cosmology,''
Phys. Rev. D \textbf{48} (1993), 4613-4622
doi:10.1103/PhysRevD.48.4613 [arXiv:astro-ph/9306029 [astro-ph]].
%192 citations counted in INSPIRE as of 10 May 2022

%\cite{Boyle:2005se}
\bibitem{Boyle:2005se}
L.~A.~Boyle and P.~J.~Steinhardt,
%``Probing the early universe with inflationary gravitational waves,''
Phys. Rev. D \textbf{77} (2008), 063504
doi:10.1103/PhysRevD.77.063504 [arXiv:astro-ph/0512014
[astro-ph]].
%196 citations counted in INSPIRE as of 10 May 2022



%\cite{Zhang:2005nw}
\bibitem{Zhang:2005nw}
Y.~Zhang, Y.~Yuan, W.~Zhao and Y.~T.~Chen,
%``Relic gravitational waves in the accelerating Universe,''
Class. Quant. Grav. \textbf{22} (2005), 1383-1394
doi:10.1088/0264-9381/22/7/011 [arXiv:astro-ph/0501329
[astro-ph]].
%72 citations counted in INSPIRE as of 10 May 2022



%\cite{Caprini:2018mtu}
\bibitem{Caprini:2018mtu}
C.~Caprini and D.~G.~Figueroa,
%``Cosmological Backgrounds of Gravitational Waves,''
Class. Quant. Grav. \textbf{35} (2018) no.16, 163001
doi:10.1088/1361-6382/aac608 [arXiv:1801.04268 [astro-ph.CO]].
%389 citations counted in INSPIRE as of 10 May 2022




%\cite{Clarke:2020bil}
\bibitem{Clarke:2020bil}
T.~J.~Clarke, E.~J.~Copeland and A.~Moss,
%``Constraints on primordial gravitational waves from the Cosmic Microwave Background,''
JCAP \textbf{10} (2020), 002 doi:10.1088/1475-7516/2020/10/002
[arXiv:2004.11396 [astro-ph.CO]].
%24 citations counted in INSPIRE as of 10 May 2022



%\cite{Smith:2005mm}
\bibitem{Smith:2005mm}
T.~L.~Smith, M.~Kamionkowski and A.~Cooray,
%``Direct detection of the inflationary gravitational wave background,''
Phys. Rev. D \textbf{73} (2006), 023504
doi:10.1103/PhysRevD.73.023504 [arXiv:astro-ph/0506422
[astro-ph]].
%193 citations counted in INSPIRE as of 10 May 2022




%\cite{Giovannini:2008tm}
\bibitem{Giovannini:2008tm}
M.~Giovannini,
%``Thermal history of the plasma and high-frequency gravitons,''
Class. Quant. Grav. \textbf{26} (2009), 045004
doi:10.1088/0264-9381/26/4/045004 [arXiv:0807.4317 [astro-ph]].
%51 citations counted in INSPIRE as of 10 May 2022

%\cite{Liu:2015psa}
\bibitem{Liu:2015psa}
X.~J.~Liu, W.~Zhao, Y.~Zhang and Z.~H.~Zhu,
%``Detecting Relic Gravitational Waves by Pulsar Timing Arrays: Effects of Cosmic Phase Transitions and Relativistic Free-Streaming Gases,''
Phys. Rev. D \textbf{93} (2016) no.2, 024031
doi:10.1103/PhysRevD.93.024031 [arXiv:1509.03524 [astro-ph.CO]].
%31 citations counted in INSPIRE as of 10 May 2022


%\cite{Giovannini:2023itq}
\bibitem{Giovannini:2023itq}
M.~Giovannini,
%``Relic gravitons and high-frequency detectors,''
[arXiv:2303.11928 [gr-qc]].
%0 citations counted in INSPIRE as of 25 Mar 2023






%\cite{Giovannini:2022eue}
\bibitem{Giovannini:2022eue}
M.~Giovannini,
%``Inflation, space-borne interferometers and the expansion history of the Universe,''
Eur. Phys. J. C \textbf{82} (2022) no.9, 828
doi:10.1140/epjc/s10052-022-10800-4 [arXiv:2206.08217 [gr-qc]].
%1 citations counted in INSPIRE as of 25 Mar 2023


%\cite{Giovannini:2022vha}
\bibitem{Giovannini:2022vha}
M.~Giovannini,
%``Relic gravitons at intermediate frequencies and the expansion history of the Universe,''
Phys. Rev. D \textbf{105} (2022) no.10, 103524
doi:10.1103/PhysRevD.105.103524 [arXiv:2203.13586 [gr-qc]].
%1 citations counted in INSPIRE as of 25 Mar 2023


%\cite{Giovannini:2020wrx}
\bibitem{Giovannini:2020wrx}
M.~Giovannini,
%``Relic gravitons from stiff curvature perturbations,''
Phys. Lett. B \textbf{810} (2020), 135801
doi:10.1016/j.physletb.2020.135801 [arXiv:2006.02760 [gr-qc]].
%2 citations counted in INSPIRE as of 25 Mar 2023


%\cite{Giovannini:2019oii}
\bibitem{Giovannini:2019oii}
M.~Giovannini,
%``Primordial backgrounds of relic gravitons,''
Prog. Part. Nucl. Phys. \textbf{112} (2020), 103774
doi:10.1016/j.ppnp.2020.103774 [arXiv:1912.07065 [astro-ph.CO]].
%23 citations counted in INSPIRE as of 25 Mar 2023


%\cite{Giovannini:2019ioo}
\bibitem{Giovannini:2019ioo}
M.~Giovannini,
%``Effective energy density of relic gravitons,''
Phys. Rev. D \textbf{100} (2019) no.8, 083531
doi:10.1103/PhysRevD.100.083531 [arXiv:1908.09679 [hep-th]].
%5 citations counted in INSPIRE as of 25 Mar 2023



%\cite{Giovannini:2014vya}
\bibitem{Giovannini:2014vya}
M.~Giovannini,
%``Scalar modes of the relic gravitons,''
Phys. Rev. D \textbf{91} (2015) no.2, 023521
doi:10.1103/PhysRevD.91.023521 [arXiv:1410.5307 [hep-th]].
%6 citations counted in INSPIRE as of 25 Mar 2023


%\cite{Giovannini:2009kg}
\bibitem{Giovannini:2009kg}
M.~Giovannini,
%``Stochastic backgrounds of relic gravitons: a theoretical appraisal,''
PMC Phys. A \textbf{4} (2010), 1 doi:10.1186/1754-0410-4-1
[arXiv:0901.3026 [astro-ph.CO]].
%45 citations counted in INSPIRE as of 25 Mar 2023



%\cite{Kamionkowski:1993fg}
\bibitem{Kamionkowski:1993fg}
M.~Kamionkowski, A.~Kosowsky and M.~S.~Turner,
%``Gravitational radiation from first order phase transitions,''
Phys. Rev. D \textbf{49} (1994), 2837-2851
doi:10.1103/PhysRevD.49.2837 [arXiv:astro-ph/9310044 [astro-ph]].
%573 citations counted in INSPIRE as of 10 May 2022

%\cite{Giare:2020vss}
\bibitem{Giare:2020vss}
W.~Giar\`e and F.~Renzi,
%``Propagating speed of primordial gravitational waves,''
Phys. Rev. D \textbf{102} (2020) no.8, 083530
doi:10.1103/PhysRevD.102.083530 [arXiv:2007.04256 [astro-ph.CO]].
%15 citations counted in INSPIRE as of 10 May 2022


%\cite{Zhao:2006mm}
\bibitem{Zhao:2006mm}
W.~Zhao and Y.~Zhang,
%``Relic gravitational waves and their detection,''
Phys. Rev. D \textbf{74} (2006), 043503
doi:10.1103/PhysRevD.74.043503 [arXiv:astro-ph/0604458
[astro-ph]].
%69 citations counted in INSPIRE as of 10 May 2022






%\cite{Lasky:2015lej}
\bibitem{Lasky:2015lej}
P.~D.~Lasky, C.~M.~F.~Mingarelli, T.~L.~Smith, J.~T.~Giblin,
D.~J.~Reardon, R.~Caldwell, M.~Bailes, N.~D.~R.~Bhat,
S.~Burke-Spolaor and W.~Coles, \textit{et al.}
%``Gravitational-wave cosmology across 29 decades in frequency,''
Phys. Rev. X \textbf{6} (2016) no.1, 011035
doi:10.1103/PhysRevX.6.011035 [arXiv:1511.05994 [astro-ph.CO]].
%147 citations counted in INSPIRE as of 10 May 2022





%\cite{Cai:2021uup}
\bibitem{Cai:2021uup}
R.~G.~Cai, C.~Fu and W.~W.~Yu,
%``Parity violation in stochastic gravitational wave background from inflation,''
[arXiv:2112.04794 [astro-ph.CO]].
%8 citations counted in INSPIRE as of 10 May 2022


%\cite{Odintsov:2021kup}
\bibitem{Odintsov:2021kup}
S.~D.~Odintsov, V.~K.~Oikonomou and F.~P.~Fronimos,
%``Quantitative predictions for f(R) gravity primordial gravitational waves,''
Phys. Dark Univ. \textbf{35} (2022), 100950
doi:10.1016/j.dark.2022.100950 [arXiv:2108.11231 [gr-qc]].
%10 citations counted in INSPIRE as of 10 May 2022






%\cite{Lin:2021vwc}
\bibitem{Lin:2021vwc}
J.~Lin, S.~Gao, Y.~Gong, Y.~Lu, Z.~Wang and F.~Zhang,
%``Primordial black holes and scalar induced secondary gravitational waves from Higgs inflation with non-canonical kinetic term,''
[arXiv:2111.01362 [gr-qc]].
%8 citations counted in INSPIRE as of 10 May 2022

%\cite{Zhang:2021vak}
\bibitem{Zhang:2021vak}
F.~Zhang, J.~Lin and Y.~Lu,
%``Double-peaked inflation model: Scalar induced gravitational waves and primordial-black-hole suppression from primordial non-Gaussianity,''
Phys. Rev. D \textbf{104} (2021) no.6, 063515 [erratum: Phys. Rev.
D \textbf{104} (2021) no.12, 129902]
doi:10.1103/PhysRevD.104.063515 [arXiv:2106.10792 [gr-qc]].
%11 citations counted in INSPIRE as of 10 May 2022

%\cite{Visinelli:2017bny}
\bibitem{Visinelli:2017bny}
L.~Visinelli, N.~Bolis and S.~Vagnozzi,
%``Brane-world extra dimensions in light of GW170817,''
Phys. Rev. D \textbf{97} (2018) no.6, 064039
doi:10.1103/PhysRevD.97.064039 [arXiv:1711.06628 [gr-qc]].
%105 citations counted in INSPIRE as of 08 Jun 2023




%\cite{Pritchard:2004qp}
\bibitem{Pritchard:2004qp}
J.~R.~Pritchard and M.~Kamionkowski,
%``Cosmic microwave background fluctuations from gravitational waves: An Analytic approach,''
Annals Phys. \textbf{318} (2005), 2-36
doi:10.1016/j.aop.2005.03.005 [arXiv:astro-ph/0412581 [astro-ph]].
%150 citations counted in INSPIRE as of 10 May 2022

%\cite{Khoze:2022nyt}
\bibitem{Khoze:2022nyt}
V.~V.~Khoze and D.~L.~Milne,
%``Gravitational waves and dark matter from classical scale invariance,''
[arXiv:2212.04784 [hep-ph]].
%0 citations counted in INSPIRE as of 01 Jan 2023


%\cite{Casalino:2018tcd}
\bibitem{Casalino:2018tcd}
A.~Casalino, M.~Rinaldi, L.~Sebastiani and S.~Vagnozzi,
%``Mimicking dark matter and dark energy in a mimetic model compatible with GW170817,''
Phys. Dark Univ. \textbf{22} (2018), 108
doi:10.1016/j.dark.2018.10.001 [arXiv:1803.02620 [gr-qc]].
%73 citations counted in INSPIRE as of 08 Jun 2023




%\cite{Oikonomou:2022xoq}
\bibitem{Oikonomou:2022xoq}
V.~K.~Oikonomou,
%``Primordial gravitational waves predictions for GW170817-compatible Einstein\textendash{}Gauss\textendash{}Bonnet theory,''
Astropart. Phys. \textbf{141} (2022), 102718
doi:10.1016/j.astropartphys.2022.102718 [arXiv:2204.06304
[gr-qc]].
%0 citations counted in INSPIRE as of 10 May 2022



%\cite{Casalino:2018wnc}
\bibitem{Casalino:2018wnc}
A.~Casalino, M.~Rinaldi, L.~Sebastiani and S.~Vagnozzi,
%``Alive and well: mimetic gravity and a higher-order extension in light of GW170817,''
Class. Quant. Grav. \textbf{36} (2019) no.1, 017001
doi:10.1088/1361-6382/aaf1fd [arXiv:1811.06830 [gr-qc]].
%74 citations counted in INSPIRE as of 08 Jun 2023




%\cite{ElBourakadi:2022anr}
\bibitem{ElBourakadi:2022anr}
K.~El Bourakadi, B.~Asfour, Z.~Sakhi, Z.~M.~Bennai and T.~Ouali,
%``Primordial black holes and gravitational waves in teleparallel Gravity,''
Eur. Phys. J. C \textbf{82} (2022) no.9, 792
doi:10.1140/epjc/s10052-022-10762-7 [arXiv:2209.08585 [gr-qc]].
%0 citations counted in INSPIRE as of 26 Sep 2022



%\cite{Sturani:2021ucg}
\bibitem{Sturani:2021ucg}
R.~Sturani,
%``Fundamental Gravity and Gravitational Waves,''
Symmetry \textbf{13} (2021) no.12, 2384 doi:10.3390/sym13122384
%2 citations counted in INSPIRE as of 05 Jul 2023




%\cite{Vagnozzi:2022qmc}
\bibitem{Vagnozzi:2022qmc}
S.~Vagnozzi and A.~Loeb,
%``The Challenge of Ruling Out Inflation via the Primordial Graviton Background,''
Astrophys. J. Lett. \textbf{939} (2022) no.2, L22
doi:10.3847/2041-8213/ac9b0e [arXiv:2208.14088 [astro-ph.CO]].
%7 citations counted in INSPIRE as of 08 Jun 2023


%\cite{Arapoglu:2022vbf}
\bibitem{Arapoglu:2022vbf}
A.~S.~Arapo\u{g}lu and A.~E.~Y\"ukselci,
%``The Effect of Nonminimally Coupled Scalar Field on Gravitational Waves from First-order Vacuum Phase Transitions,''
[arXiv:2210.16699 [gr-qc]].
%0 citations counted in INSPIRE as of 02 Jan 2023


%\cite{Giare:2022wxq}
\bibitem{Giare:2022wxq}
W.~Giar\`e, M.~Forconi, E.~Di Valentino and A.~Melchiorri,
%``Towards a reliable calculation of relic radiation from primordial gravitational waves,''
[arXiv:2210.14159 [astro-ph.CO]].
%1 citations counted in INSPIRE as of 02 Jan 2023


%\cite{Oikonomou:2021kql}
\bibitem{Oikonomou:2021kql}
V.~K.~Oikonomou,
%``A refined Einstein\textendash{}Gauss\textendash{}Bonnet inflationary theoretical framework,''
Class. Quant. Grav. \textbf{38} (2021) no.19, 195025
doi:10.1088/1361-6382/ac2168 [arXiv:2108.10460 [gr-qc]].
%27 citations counted in INSPIRE as of 02 Jan 2023


%\cite{Gerbino:2016sgw}
\bibitem{Gerbino:2016sgw}
M.~Gerbino, K.~Freese, S.~Vagnozzi, M.~Lattanzi, O.~Mena,
E.~Giusarma and S.~Ho,
%``Impact of neutrino properties on the estimation of inflationary parameters from current and future observations,''
Phys. Rev. D \textbf{95} (2017) no.4, 043512
doi:10.1103/PhysRevD.95.043512 [arXiv:1610.08830 [astro-ph.CO]].
%72 citations counted in INSPIRE as of 04 Jul 2023
%\cite{Vagnozzi:2022moj}

%\cite{Breitbach:2018ddu}
\bibitem{Breitbach:2018ddu}
M.~Breitbach, J.~Kopp, E.~Madge, T.~Opferkuch and P.~Schwaller,
%``Dark, Cold, and Noisy: Constraining Secluded Hidden Sectors with Gravitational Waves,''
JCAP \textbf{07} (2019), 007 doi:10.1088/1475-7516/2019/07/007
[arXiv:1811.11175 [hep-ph]].
%147 citations counted in INSPIRE as of 04 Jul 2023




%\cite{Pi:2019ihn}
\bibitem{Pi:2019ihn}
S.~Pi, M.~Sasaki and Y.~l.~Zhang,
%``Primordial Tensor Perturbation in Double Inflationary Scenario with a Break,''
JCAP \textbf{06} (2019), 049 doi:10.1088/1475-7516/2019/06/049
[arXiv:1904.06304 [gr-qc]].
%20 citations counted in INSPIRE as of 13 Jul 2023






%\cite{Khlopov:2023mpo}
\bibitem{Khlopov:2023mpo}
M.~Khlopov and S.~R.~Chowdhury,
%``Polarization of Gravitational Waves in Modified Gravity,''
Symmetry \textbf{15} (2023) no.4, 832 doi:10.3390/sym15040832
%0 citations counted in INSPIRE as of 05 Jul 2023


%\cite{Odintsov:2022cbm}
\bibitem{Odintsov:2022cbm}
S.~D.~Odintsov, V.~K.~Oikonomou and R.~Myrzakulov,
%``Spectrum of Primordial Gravitational Waves in Modified Gravities: A Short Overview,''
Symmetry \textbf{14} (2022) no.4, 729 doi:10.3390/sym14040729
[arXiv:2204.00876 [gr-qc]].
%0 citations counted in INSPIRE as of 10 May 2022








%\cite{Watanabe:2006qe}
\bibitem{Watanabe:2006qe}
Y.~Watanabe and E.~Komatsu,
%``Improved Calculation of the Primordial Gravitational Wave Spectrum in the Standard Model,''
Phys. Rev. D \textbf{73} (2006), 123515
doi:10.1103/PhysRevD.73.123515 [arXiv:astro-ph/0604176
[astro-ph]].
%234 citations counted in INSPIRE as of 06 Sep 2023



%\cite{Kuroyanagi:2014nba}
\bibitem{Kuroyanagi:2014nba}
S.~Kuroyanagi, T.~Takahashi and S.~Yokoyama,
%``Blue-tilted Tensor Spectrum and Thermal History of the Universe,''
JCAP \textbf{02} (2015), 003 doi:10.1088/1475-7516/2015/02/003
[arXiv:1407.4785 [astro-ph.CO]].
%81 citations counted in INSPIRE as of 06 Sep 2023



%\cite{Kamionkowski:2022pkx}
\bibitem{Kamionkowski:2022pkx}
M.~Kamionkowski and A.~G.~Riess,
%``The Hubble Tension and Early Dark Energy,''
[arXiv:2211.04492 [astro-ph.CO]].
%88 citations counted in INSPIRE as of 28 Oct 2023



\end{thebibliography}
\end{document}